\documentclass[12pt]{article}
\usepackage{scicite}
\usepackage{times}
\usepackage[T1]{fontenc}
\usepackage{float}
\usepackage{graphicx}
\usepackage{pdflscape}
\usepackage{hyperref}
\usepackage{pgf,pgffor}
\usepackage{textcomp}
\usepackage{xcolor}
\usepackage{ulem}
\usepackage{lineno}
\usepackage{caption}
\usepackage[caption = false]{subfig}
\usepackage{pifont}
\usepackage{longtable}
\usepackage{amssymb}
\usepackage{cleveref}
\definecolor{orcidlogocol}{HTML}{A6CE39}
\usepackage[symbol]{footmisc}

\usepackage{deluxetable}
\newcommand{\HI}{{H\,\textsc{i}}}

\newcommand{\kms}{km~s$^{-1}$}

\newcommand{\bain}{Bulletin of the Astronomical Institutes of the Netherlands}
\newcommand{\aj}{Astron. J.}
\newcommand{\apj}{Astrophys. J.}
\newcommand{\apjl}{Astrophys. J. Lett.}
\newcommand{\apjss}{Astrophys. J. Suppl. Ser.}
\newcommand{\apjs}{Astrophys. J. Suppl. }
\newcommand{\aap}{Astron. Astrophys.}
\newcommand{\mnras}{Mon. Not. R. Astron. Soc.}
\newcommand{\nat}{Nature}
\newcommand{\araa}{Ann. Rev. Astron. Astrophys.}

\makeatletter

\let\saved@includegraphics\includegraphics
\AtBeginDocument{\let\includegraphics\saved@includegraphics}
\renewenvironment*{figure}{\@float{figure}}{\end@float}
\makeatother

\makeatletter
\def\@fnsymbol#1{\ensuremath{\ifcase#1\or \dagger\or \ddagger\or
 \mathsection\or \mathparagraph\or \|\or **\or \dagger\dagger
 \or \ddagger\ddagger \else\@ctrerr\fi}}
\makeatother

\newcommand{\AFF}[1]{$^{\foreach\d[count=\ni]in{#1}{\ifnum\ni=1\ref{\d}\else,\ref{\d}\fi}}$}
\newenvironment{sciabstract}{%
\begin{quote} \bf}
{\end{quote}}

\title{Discovery of a high-velocity cloud of the Milky Way as a potential dark galaxy}
\author{Xiao-Lan Liu$^{\rm 1,2}$, Jin-Long Xu$^{\rm 1,2^\ast}$, Peng Jiang$^{\rm 1,2^\ast}$, Ming Zhu$^{\rm 1,2^\ast}$, \\
 Chuan-Peng Zhang$^{\rm 1,2}$, Naiping Yu$^{\rm 1,2}$, Ye Xu$^{\rm 3}$, Xin Guan$^{\rm 1,2}$, Jun-Jie Wang$^{\rm 1,2}$ \\
\footnotesize{$^1$National Astronomical Observatories, Chinese Academy of Science, Beijing, 100101, China.}\\ 
\footnotesize{$^2$Guizhou Radio Astronomical Observatory, Guizhou University, Guiyang 550000, China.} \\
\footnotesize{$^3$Purple Mountain Observatory, Chinese Academy of Sciences, Nanjing 210008, China.} \\
\footnotesize{Corresponding author. Email:xujl@bao.ac.cn; pjiang@bao.ac.cn; mz@bao.ac.cn }
}

\date{}

\begin{document}
\baselineskip24pt
\maketitle

\begin{sciabstract}
High-velocity clouds (HVCs) are composed of neutral hydrogen (\HI) moving at velocities that deviate from the general rotational motion of the Milky Way. Currently, the origins of the HVCs remain poorly known due to the difficulty in determining their distance and the lack of any other suitable identification. Here we report the detection of a compact gas clump in HVC AC-I, which displays characteristics typical of a disk galaxy, named AC G185.0-11.5, using the \HI\ observations. We estimated the distance of AC G185.0-11.5 to be 277.7$^{+291.3}_{-141.6}$  kpc using the Baryonic Tully-Fisher relation and constrained its \HI\ gas mass to be between 3.0$\times10^7$ and 4.7$\times10^8$ solar masses. The distance determination indicates that the HVC AC-I hosting AC G185.0-11.5 is an extragalactic object in the Local Group. The absence of molecular gas and an optical counterpart for AC G185.0-11.5 implies that it may be a rare dark galaxy.  
\end{sciabstract}

\section*{Introduction}
\vspace{3mm}
High-velocity clouds (HVCs) are defined by their line-of-sight velocities exceeding 90 km s$^{-1}$ relative to the local standard of rest (LSR)\cite{Wakker+1997}. Numerous HVCs have been detected near our Milky Way (MW)\cite{Wakker+1991,Putman+2012,Westmeier+2018}. These clouds play a critical role in regulating the gaseous environment and star formation rate of the MW\cite{Putman+2012,Richter+2017,Fox+2019}. Three primary formation scenarios for HVCs have been proposed: gas ejected from the MW disk by supernovae, remnants from the early galaxy formation era possibly accreted into the MW halo, or genuine extragalactic objects within the Local Group (LG) of galaxies\cite{Woerden+1999+nature, Lucchini+2024}.  An important challenge in understanding HVC origins stems from their unknown distances. Since HVCs lack objects of fixed luminosity and do not conform to Galactic rotation, distance measurements rely on optical and UV absorption lines from the chance position of stars\cite{Wakker+1997}. Usually, there are no bright stars located behind a given cloud.  Therefore, searching for various possible tracers of the HVCs will play a crucial role in revealing their origin.

Complex AC-I is one part of the Galactic anticenter HVCs, originally identified as a high negative-velocity hydrogen cloud in surveys\cite{Mathewson+1966, Hulsbosch+1968}. AC-I spans velocities from $-230.0$ \kms\ to $-168.0$ km s$^{-1}$ \cite{Tamanaha+1997}. A search for star absorption lines towards AC-I revealed no associated high-velocity interstellar absorption, setting a lower distance limit of approximately 650 pc\cite{Tamanaha+1996}. AC-I seems to show a consistent variation in radial velocity along its length, suggesting possible rotational motion. To maintain stability, AC-I would need to be at a distance greater than 500 kpc, implying it is likely an extragalactic object based on its distance\cite{Verschuur+1969}. However, the current limited evidence is insufficient to determine the origin of HVC AC-I. Figure~\ref{contourMaps}a displays the \HI\ column density map of AC-I from the Galactic Arecibo L-Band Feed Array \HI\ (GALFA-\HI) survey data. The \HI\ emission reveals an extended structure extending from southeast to northwest, consistent with previous low-resolution observations of its radial velocity variation\cite{Tamanaha+1997}. Higher-resolution analysis shows that AC-I can be decomposed into a gas clump and surrounding gas resembling tails, as a head-tail structure\cite{Meyerdierks+1991, Bruns+2000}, rather than a complete rotational structure as previously proposed\cite{Verschuur+1969}.

\section*{Results and Discussion}
We conducted new higher-resolution \HI\ observations of the gas clump with relatively high column density within HVC AC-I using the Five-hundred-meter Aperture Spherical radio Telescope (FAST). In Fig.~\ref{contourMaps}b, the gas clump exhibits a compact spherical-like structure, featuring a hollow on its northeastern edge and a gas concentration towards its center. Previous studies reported column densities for compact HVCs ranging from 2.0$\times10^{18}$ to 3.4$\times10^{19}$ cm$^{-2}$ with line widths between 6.3 and 34.3 km s$^{-1}$ \cite{Saul+2012}. In our measurements, we found that the peak column density in the compact clump of AC-I is approximately 3.8$\times10^{20}$ cm$^{-2}$, exceeding those of typical compact HVCs by at least one order of magnitude but comparable to faint gas-rich dwarf galaxies Leo T and Leo P in the LG\cite{Ryan-Weber+2008, Giovanelli+2013}. The global \HI\ profile of the compact clump is displayed in Fig.~\ref{integratedSpectrum}, for which we performed Gaussian fitting. The derived system velocity ($V_{\rm LSR}$), line width at 50$\%$ of the peak flux ($W_{50}$), and total flux ($S_{\rm t}$) are -193.0$^{+0.2}_{-0.2}$ km s$^{-1}$, 39.6$^{+0.5}_{-0.5}$ km s$^{-1}$, and 6.2$^{+0.1}_{-0.1}$ $\times10^{3}$ Jy km s$^{-1}$, respectively. The $S_{\rm t}$ value of this clump is two orders of magnitude lower than those observed in the Large and Small Magellanic Clouds\cite{Bruns+2005}, yet substantially higher than other compact HVCs\cite{Saul+2012, Adams+2013} and known LG dwarf galaxies\cite{Ryan-Weber+2008, Giovanelli+2013}. The larger $\vert V_{\rm LSR}\vert$, higher column density, and abnormally large total flux indicate that this compact clump is distinct from typical HVCs.

The velocity-field map in Fig.~\ref{model}a shows a consistent velocity gradient along the major axis of the special gas clump, as indicated by the blue arrow in Fig. \ref{contourMaps}a.  Several dynamic processes can explain the velocity gradients observed in HVCs. Due to the galactic fountain driven by supernovae failing to explain \HI\ clouds at velocities exceeding 100 \kms\ from a galaxy's disk \cite{Blitz+1999, van+1988, Marasco+2012}, we can first rule out the possibility that the velocity gradient in the clump arises from such a dynamic process. Moreover, both the ram pressure when an HVC falls into a massive galaxy or galaxy group and the interaction with the surrounding gas of the Galaxy halo can cause the observed velocity gradients \cite{Peek+2007, Lockman+2008}. Such HVCs generally display the head-tail structure. We have indeed observed that the HVC AC-I shows a head-tail structure with its direction running from northwest to southeast, as indicated by the red arrow in Fig. \ref{contourMaps}a. However, the velocity gradient observed in the compact clump is nearly perpendicular to both the head-tail structure and the velocity variation in the HVC AC-I \cite{Tamanaha+1997}. This suggests that the velocity gradient in the compact clump does not reflect the motion of the HVC AC-I and is unlikely to result from these dynamic processes. Instead, such a velocity gradient is more consistent with the presence of a rotating gas disk within the compact clump.

To further reveal the kinematic properties of this compact clump, we generated the velocity channel maps spanning from -238.4 km s$^{-1}$ to -141.8 km s$^{-1}$, depicted in {Fig.~\ref{channelGalaxy}. These maps display that the \HI\ gas emission changes with the velocity along the major axis of the clump, which runs from southwest to northeast. The gas with the largest redshift is found at one side of the major axis, but as the velocity becomes less red the emission moves to the other side of this axis, further indicative of a rotating disk structure \cite{Genzel+etal+2006+nature}. The presence of a rotating gas disk is akin to a protostar with a Keplerian disk \cite{Lu+etal+2022+NA} or a disk galaxy. Figure~\ref{model}b shows the position-velocity (PV) diagram cutting along the major axis of this clump. The PV diagram displays a perfect S-like structure, not a Keplerian rotation shape. The S-like structure is considered as the most typical characteristic of disk galaxies\cite{Neeleman+etal+2020+nature, Rizzo+etal+2020+nature}, yet no such structure has been known in HVCs. Based on the multiple evidence mentioned above, we conclude that the compact clump in AC-I is a disk galaxy with very high confidence. We have designated this galaxy as AC G185.0-11.5 due to its association with the HVC AC-I. 

The typical method for estimating distances to HVCs involves using stars with known distances \cite{Woerden+1999+nature,Thom+2008}. However, previous observations with the Hubble Space Telescope did not detect any HVCs with $\vert V_{\rm LSR}\vert > 170$ \kms\ (sometimes called very high-velocity clouds, VHVC) in absorption in stellar spectra \cite{Lehner+2011+sci,Lehner+2022}. These findings led to the conclusion that such VHVCs must be located at distances greater than 10–15 kpc. AC G185.0-11.5 has a systemic velocity of $\vert V_{\rm LSR}\vert=$ $193.0_{-0.2}^{+0.2}$ \kms, implying that it is a VHVC and its distance should be at least greater than 10 kpc. To further constrain the distance to AC G185.0-11.5, we identified the most distant star (9.9 kpc) projected onto the HVC AC-I from the Gaia DR3 catalog \cite{Collaboration+2023}. No absorption was detected in the optical spectrum of this star at the system velocity of the HVC AC-I (see Materials and Methods for details). This observation indicates that the HVC is located behind the star, further supporting the conclusion that the distance to AC G185.0-11.5 within HVC AC-I is greater than 10 kpc.

The Tully–Fisher (TF) relation has proven effective in determining cosmic distances\cite{Tully-Fisher+1977, Makarov+2014}. A related empirical correlation, the Baryonic Tully-Fisher (BTF) relation, links a galaxy's rotation velocity ($V_{\rm rot}$) to its total baryonic mass ($M_{\rm bary}$)\cite{McGaugh+2000, McGaugh+2012, McGaugh+2015, Lelli+2016}, making it a robust tool for disk galaxies. The BTF relation is similar to the TF relation and has successfully constrained the distance of the well-known dwarf galaxy Leo P\cite{Giovanelli+2013}. Given that AC G185.0-11.5 exhibits rotational characteristics typical of disk galaxies, we can apply the BTF relation to estimate its distance. Figure~\ref{distance}a illustrates the position of AC G185.0-11.5 on the BTF relation. From this, we calculated $M_{\rm bary}$ for AC G185.0-11.5 as $1.5^{+4.8}_{-1.1}\times10^8\rm\, M_\odot$ based on its $V_{\rm rot}$ (see Materials and Methods for details). Using the total flux of the galaxy, we derived a distance estimate of 277.7$^{+291.3}_{-141.6}$ kpc, as shown in Fig.~\ref{distance}b, placing AC G185.0-11.5 within the LG. Based on statistical and theoretical arguments regarding the O VI (O$^{5+}$) absorbers detected in the ultraviolet emission from active galactic nuclei, compact HVCs have been proposed to originate at large characteristic distances  ($>$100 kpc) in the LG \cite{Nicastro+2003+NA}. AC G185.0-11.5, as a compact HVC, is fully consistent with these previous findings.

Star formation in galaxies often correlates with regions of high column density. Simulations and observations indicate that when the \HI\ column density exceeds 5.0$\times10^{20}$ cm$^{-2}$, galaxy disks can cool and form molecular hydrogen (H$_{2}$), facilitating subsequent star formation\cite{Davies+2006}. We measured a peak column density of approximately 3.8$\times10^{20}$ cm$^{-2}$, slightly below the threshold required for H$_{2}$ formation and subsequent star formation. Observing H$_{2}$ directly is challenging, so carbon monoxide (CO) is often used as a tracer for H$_{2}$ emission. We conducted CO ($J=1-0$) observations toward the peak position of AC G185.0-11.5 but did not detect any CO emission, with a 3$\sigma$ sensitivity limit of 0.03 K \kms\ per channel. Using the ratio between H$_{2}$ column density and integrated CO intensity for high-latitude translucent clouds\cite{Magnani-Onello+1995}, $X$(CO)=1.2$\times10^{20}$ cm$^{-2}$ (K \kms)$^{-1}$, we estimate an H$_{2}$ column density of 3.6$\times10^{18}$ cm$^{-2}$. This suggests that the H$_{2}$ column density in AC G185.0-11.5 is lower than 3.6$\times10^{18}$ cm$^{-2}$, consistent with the lower \HI\ column density observed compared to the star formation threshold.

To identify potential stellar components associated with AC G185.0-11.5, we utilized the Pan-STARRS1 $g$-band image\cite{Chambers+2016}. Figure~\ref{contourMaps}c presents the Pan-STARRS1 $g$-band mosaic image, focusing on the denser region of AC G185.0-11.5. No extended optical emission is evident in the dense region of this galaxy. To further check whether the stellar populations are overdensity in the dense region, we used the Gaia DR3 catalog\cite{Collaboration+2023}. Regarding the distribution of the selected stars, we plot the map of the star surface density, as shown in Fig.~\ref{Gaiaoverdensity}. We still haven't seen clear signs of clustering toward the dense region. One factor contributing to the absence of an optical counterpart is the detection limit of the Pan-STARRS1 data used in our analysis. We applied the $g$-band TF relation, given by $M_{\rm g}=a \times \log_{10} (2V_{\rm rot}) + b$, where $M_{g}$ denotes the absolute magnitude. For galaxies with known distances, typical coefficients\cite{Ponomareva+2017} are $a$=-7.12$\pm$0.6 and $b$=-1.94$\pm$1.51. According to this relation, AC G185.0-11.5 would have an expected $M_{g}$ of -15.7$\pm$1.9 mag. Correcting for Galactic extinction using a new three-dimensional map of dust reddening\cite{Green+19}, we estimated an apparent magnitude range of 4.3 to 11.2 mag in the $g$-band for AC G185.0-11.5, considerably‌ brighter than the detection limit (23.4 mag) of the Pan-STARRS1 survey\cite{Chambers+2016}. Therefore, an optical counterpart in AC G185.0-11.5 should be detectable in the Pan-STARRS1 survey image. The absence of molecular gas and an optical counterpart suggests that the galaxy AC G185.0-11.5 may be devoid of stars.

AC G185.0-11.5 shows a spherical-like structure with a hollow on one side and a tail-like structure towards the other, as shown in red arrows in Fig.~\ref{channelGalaxy}. The features suggest AC G185.0-11.5 has undergone a gas stripping through ram pressure or tidal interaction. Tidal dwarf galaxies typically lack dark matter ($M_{\rm dyn}/M_{\rm bar} \simeq 1$) and are often found in tidal tails\cite{Duc+04, Lelli+15, van+22}. We estimated the dynamical mass ($M_{\rm dyn}$) of AC G185.0-11.5 to be $3.2^{+3.5}_{-1.6} \times 10^9 \rm\,M_\odot$, resulting in $M_{\rm dyn}/M_{\rm bary}$ of 21.6$_{-11.1}^{+22.5}$. This ratio indicates the dominance of dark matter in this galaxy, consistent with our observation of a flat rotation curve (see Fig.~\ref{RotationCurve}). Thus, we can dismiss AC G185.0-11.5 as a tidal dwarf galaxy formed through tidal interactions. Moreover, the Reionization-limited \HI\ Clouds (RELHICS) are characterized by a nearly-spherical \HI\ core in hydrostatic and thermal equilibrium with the ionizing ultraviolet background\cite{Llambay+17}. Using the derived values of $V_{\rm rot}$ and $V_\sigma$ from the 3D-Based Analysis of Rotating Objects via Line Observations ($\rm ^{3D}BAROLO$) model, we estimated the ratio $V_{\rm rot}/V_\sigma$ to be $4.1_{-0.2}^{+0.2}$, providing evidence that AC G185.0-11.5 is rotation-dominated and not a RELHIC.

AC G185.0-11.5 exhibits features typical of a disk galaxy but lacks extended optical counterparts. We propose that the starless AC G185.0-11.5 may be a dark galaxy. Dark galaxies are hypothesized to be dark matter halos that contain a baryonic disk but are devoid of stars \cite{Davies+2006, Roman+21}. In the LG, HVC Complex H and HVC 127-41-330 have been speculated to be dark galaxies \cite{Simon+2006, Robishaw+2002}, but they lack a velocity structure consistent with a rotating or gravitationally bound disk. Additionally, determining whether these HVCs are dominated by dark matter remains challenging. AC G185.0-11.5, as an HVC, is likely the best current prototype of a dark galaxy in the LG. This could provide more accurate initial conditions and physical parameters for simulating galaxy formation in cosmology.

The distance estimate confirms that HVC AC-I, which hosts AC G185.0-11.5, is an extragalactic object within the LG. The dwarfs observed in the LG are predominantly gas-poor, whereas gas-rich dwarfs are typically located far from the Milky Way (MW) or M31 \cite{McConnachie+2012}. This observation has long been interpreted as evidence of the important role of external processes, such as tidal interactions or ram-pressure stripping, in the evolution of dwarf galaxies. HVC AC-I exhibits a head-tail structure, with its tail pointing toward the MW, indicating that it is moving away from the MW—an orientation distinct from that of the known Smith HVC \cite{Lockman+2008}. The cosmological virial radius of the MW halo is approximately 300 kpc \cite{Klypin+2002}. Hence, the possibility that the HVC AC-I has been affected by the tidal forces of the MW halo cannot be entirely ruled out. It is more plausible, however, that a dark matter halo recently entered the LG and underwent ram-pressure stripping, as demonstrated by numerical simulations \cite{Roediger+2008}. The less bound material has been stripped from the outer disc of the halo, leading to the formation of galaxy AC G185.0-11.5 and its irregular tails. According to numerical simulations \cite{Plockinger+2012}, AC G185.0-11.5 was not completely stripped because dark matter predominates.

It is well-known that the observed number density of low-mass dwarf galaxies in the LG is significantly lower than predicted by cold dark matter (CDM) simulations, known as the missing satellite problem\cite{Klypin+99, Moore+99}. Here, we have identified a low-mass halo. Discovering more such low-mass halos associated with known HVCs could potentially mitigate the missing satellite problem.

\begin{figure*}[h!]
   \centering
  \includegraphics[width=16cm, angle=0]{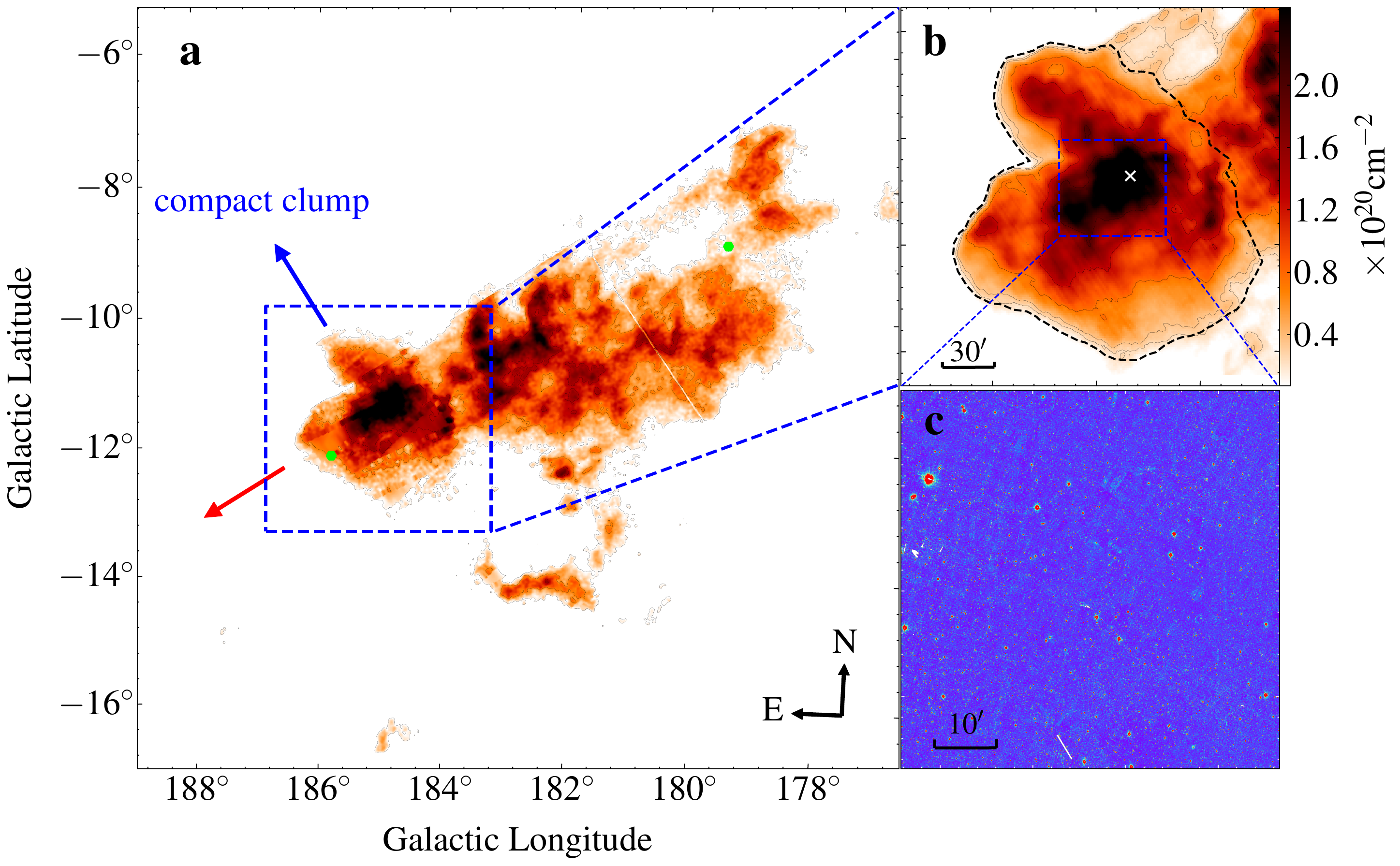}
  \vspace{-8mm}
   \caption{\textbf{Overview of the high-velocity cloud (HVC) AC-I.} \textbf{a,} \HI\ column density map across velocities from $-234.5$ km s$^{-1}$ to $-150.5$ km s$^{-1}$ using the Galactic Arecibo L-Band Feed Array \HI\ (GALFA-HI) survey data. Contours: 3.5 (5$\sigma$), 7.5, 12.5, 19.3, 24.5, and 29.6$\rm \times10^{19}\, cm^{-2}$. The red arrow indicates the motion direction of the HVC AC-I. Both a velocity gradient and the position of the compact clump (AC G185.0-11.5) are indicated by a blue arrow. The green dots indicate the positions of the selected Gaia stars. \textbf{b,} \HI\ column density map of the compact clump within AC-I, spanning velocities from -241.7 km s$^{-1}$ to -157.9 km s$^{-1}$ using the Five-hundred-meter Aperture Spherical radio Telescope (FAST) data. Contours: 2.5 (5$\sigma$), 3.5, 7.5, 12.5, 19.3, 24.5, 29.6, 37.7$\rm \times10^{19}\, cm^{-2}$. The white cross denotes the peak \HI\ column density position where CO ($J=1-0$) emission was observed. The black dashed line encircles the region integrated into the spectrum (Fig.~\ref{integratedSpectrum}). \textbf{c,} the Pan-STARRS1 $g$-band image covering the dense region of the clump. Scale bars of $30^\prime$ and 10$^\prime$ are added in panels b and c, respectively.}
   \label{contourMaps}
   \end{figure*}

\begin{figure}[h!]
\centering
\includegraphics[width=9cm, angle=0]{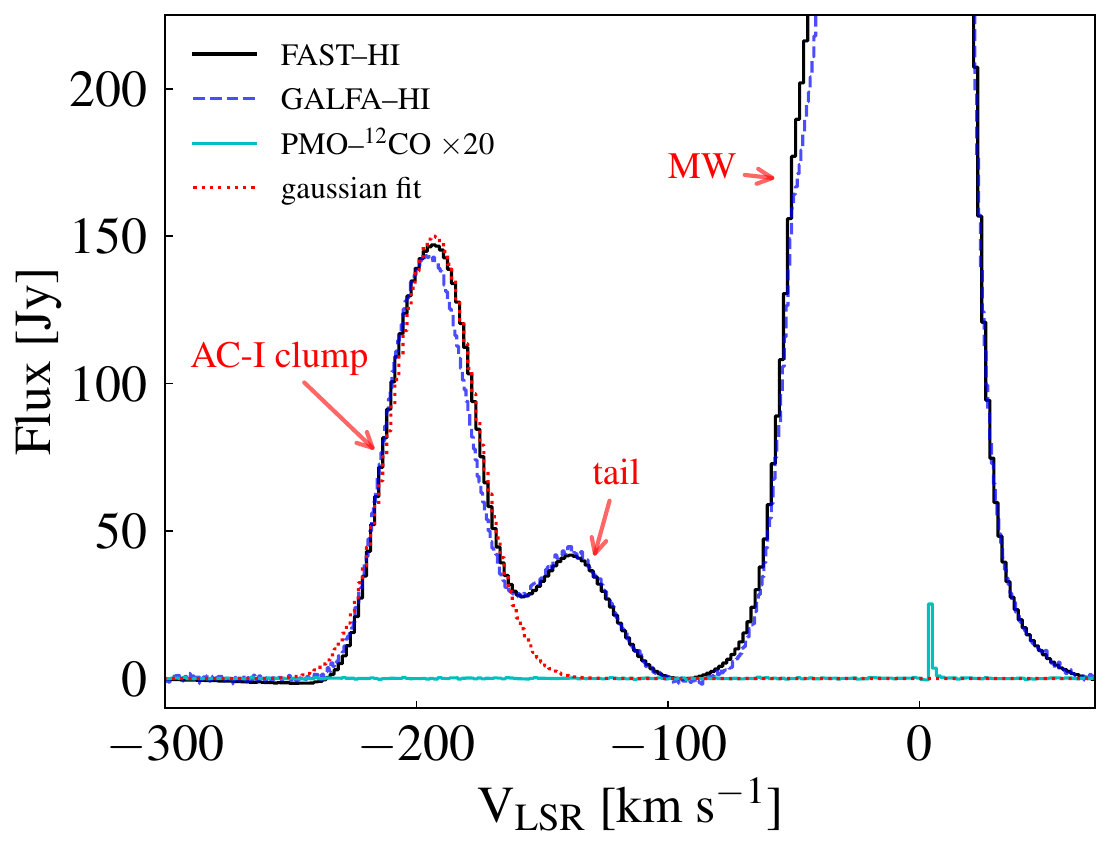}
\vspace{-3mm}
\caption{\textbf{Integrated \HI\ profile of the compact clump.}  The region integrated in the spectra is shown in panel b of Figure~\ref{contourMaps}. We only performed a Gaussian fitting on the spectrum of AC-I clump from the FAST data. The feature at -140 \kms\ indicates the partial gas component of AC-I tail.}
\label{integratedSpectrum}  
\end{figure}

\begin{figure*}[h!]
   \centering
  \includegraphics[width=16cm, angle=0]{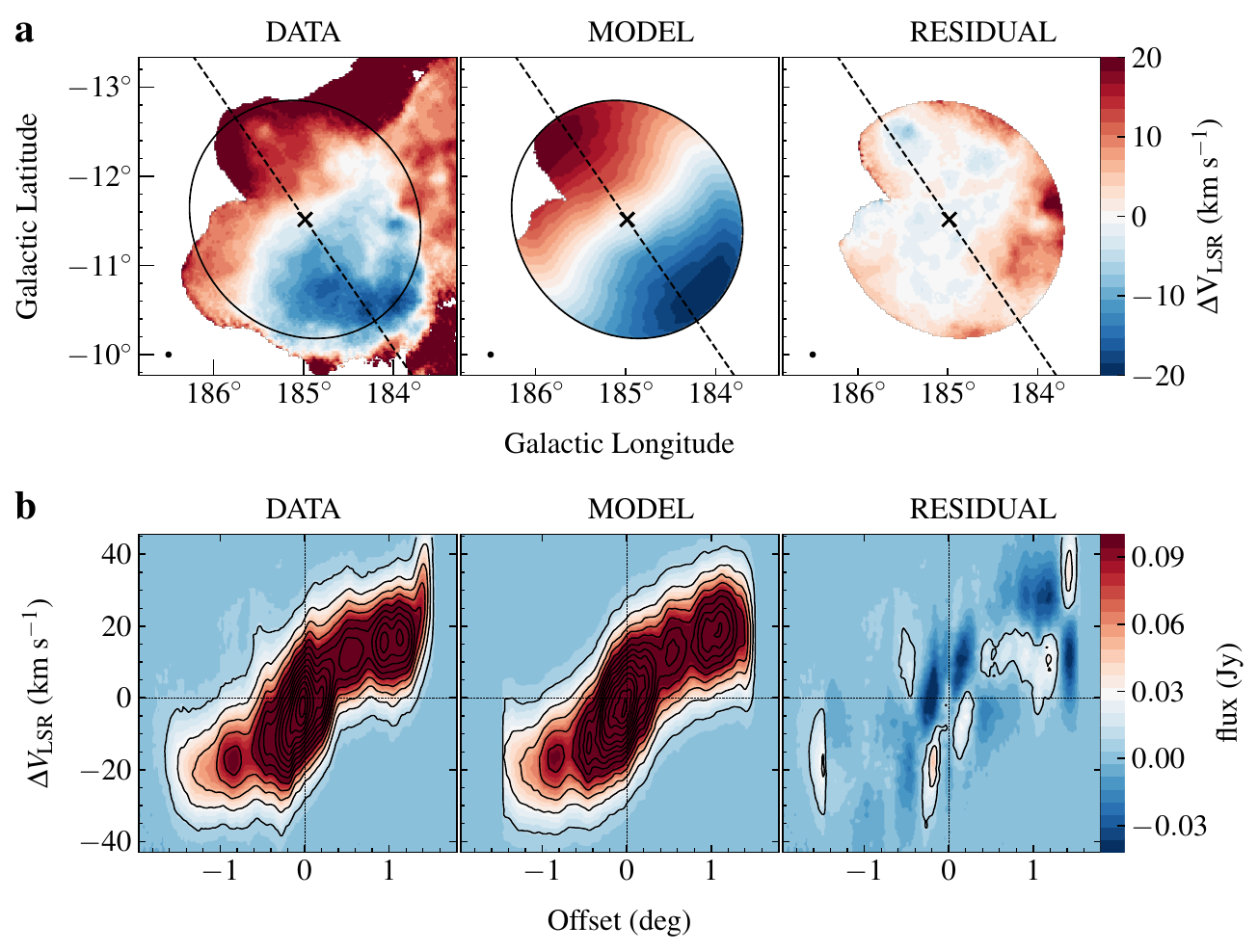}
  \vspace{-9mm}
   \caption{\textbf{Observation versus model for the compact clump.} \textbf{a,} Intensity-weighted velocity field (Moment 1) maps relative to the systemic velocity ($-193.0$ km s$^{-1}$) of the \HI\ emission for the data (left panel), the model (middle panel) and the residual (right panel). The dashed black line in each frame marks the major axis of the compact clump. The black cross denotes the kinematical center of the compact clump in each panel. The black circle manifests the outmost radius of the galaxy the model simulated. The beam size of the FAST observations is shown in the bottom-left corner. \textbf{b,} Position-Velocity (PV) map of the compact clump along the major axis for the data (left panel), the model (middle panel), and the residual (right panel). The contour levels of the data and the model start from 5$\sigma$ (1$\sigma$=0.002 Jy) to 140$\sigma$ by step of 10$\sigma$.}
   \label{model}
   \end{figure*}

\begin{figure*}[h!]
   \centering
  \includegraphics[width=15.0cm, angle=0]{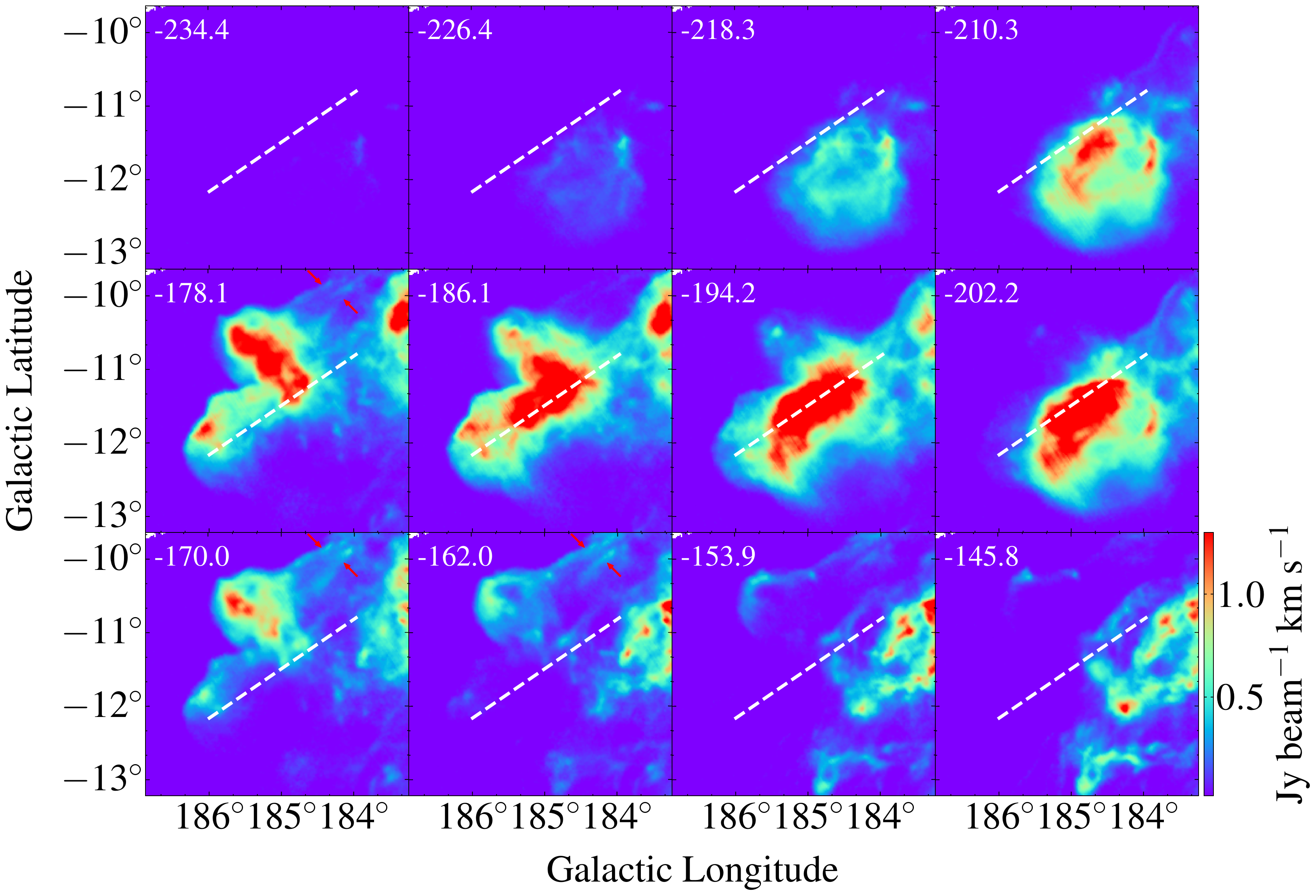}
  \vspace{-4mm}
   \caption{\textbf{Channel maps of the compact clump.} The velocity ranges from -238.4 km s$^{-1}$ to -141.8 km s$^{-1}$. A white dashed line represents the minor axis of the clump in each frame. The central velocity is shown in the top-left corner of each frame in units of km s$^{-1}$. The red arrows in panels -178.1, -170.0 and -162.0 \kms\ indicate the stripped gas.}
   \label{channelGalaxy}
   \end{figure*}
   
\begin{figure*}[h!]
  \centering
  \includegraphics[width=1.15\columnwidth]{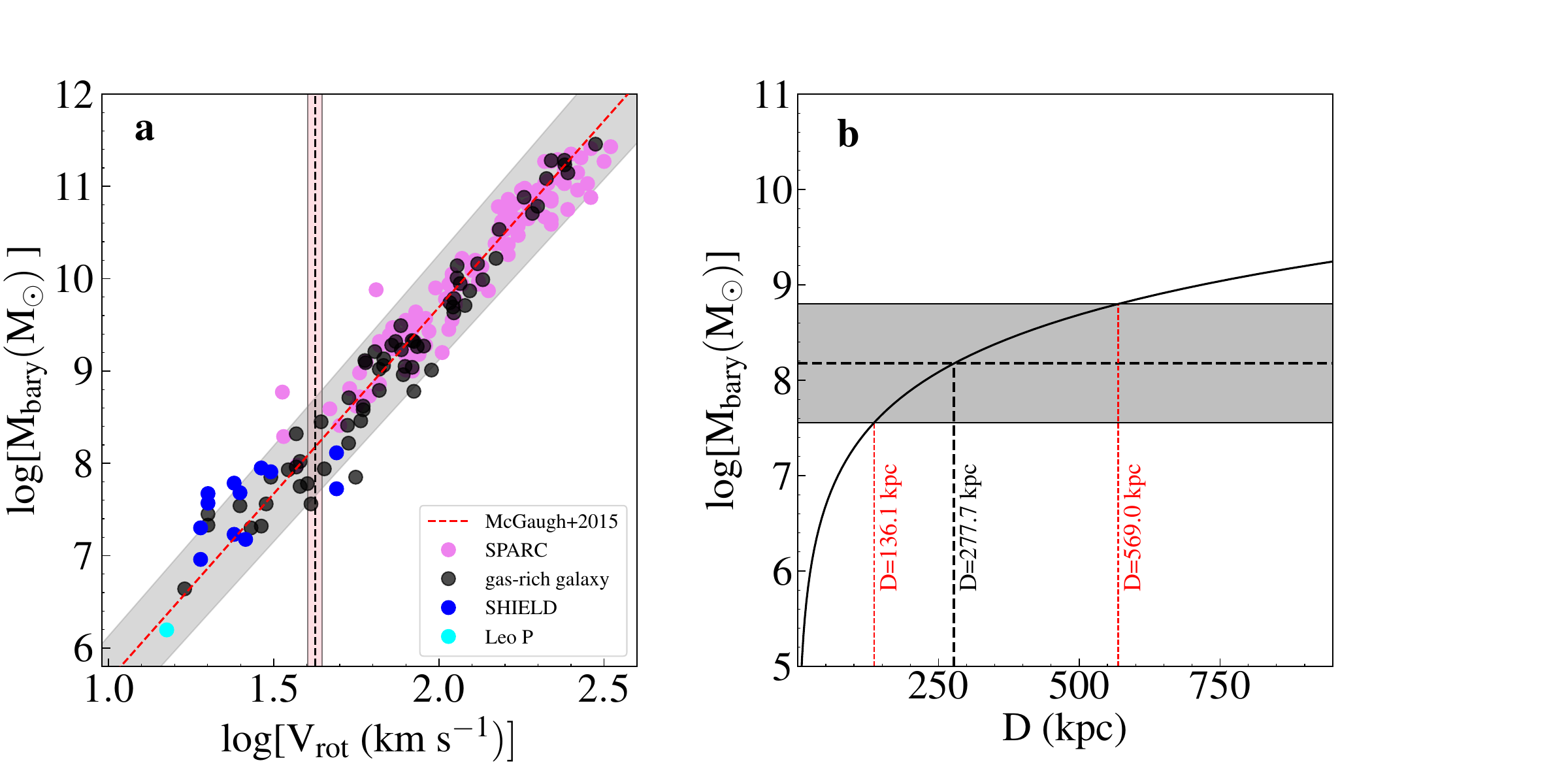}  
  \vspace{-10mm}
   \caption{\textbf{Determination of distance to galaxy AC G185.0-11.5.} \textbf{a,} baryonic Tully-Fisher (BTF) relation. The dashed red line illustrates a best-fit BTF relation for the gas-rich disk galaxies\cite{McGaugh+2015}. The gray-shaded region represents the $2\sigma$ scatter for the BTF relation. The violet-filled and blue-filled circles represent the galaxies obtained from the Spitzer Photometry and Accurate Rotation Curves (SPARC)\cite{Lelli+2016} and the Survey of \HI\ in Extremely Low-mass Dwarfs (SHIELD)\cite{McNichols+2016} samples, respectively. The black-filled circles indicate the gas-rich galaxy sample from ref.~\cite{McGaugh+2012,McGaugh+2015}. The cyan-filled circle marks the location of the dwarf galaxy Leo P\cite{Giovanelli+2013, Bernstein-Cooper+2014} on the BTF relation. The dashed black vertical line denotes the rotation velocity of AC G185.0-11.5 with its $1\sigma$ uncertainties (the pink-shaded region). \textbf{b,} Baryonic mass-distance relation. The dashed black horizontal line indicates the total baryonic mass derived from the BTF relation in panel a. The vertical black line denotes the distance to AC G185.0-11.5. The gray-shaded region illustrates the estimated baryonic mass under the $2\sigma$ scatter of the BTF relation, yielding a distance range of 136.1 kpc to 569.0 kpc for AC G185.0-11.5.} 
   \label{distance}
   \end{figure*}

\begin{figure*}[h!]
   \centering
  \includegraphics[width=10cm, angle=0]{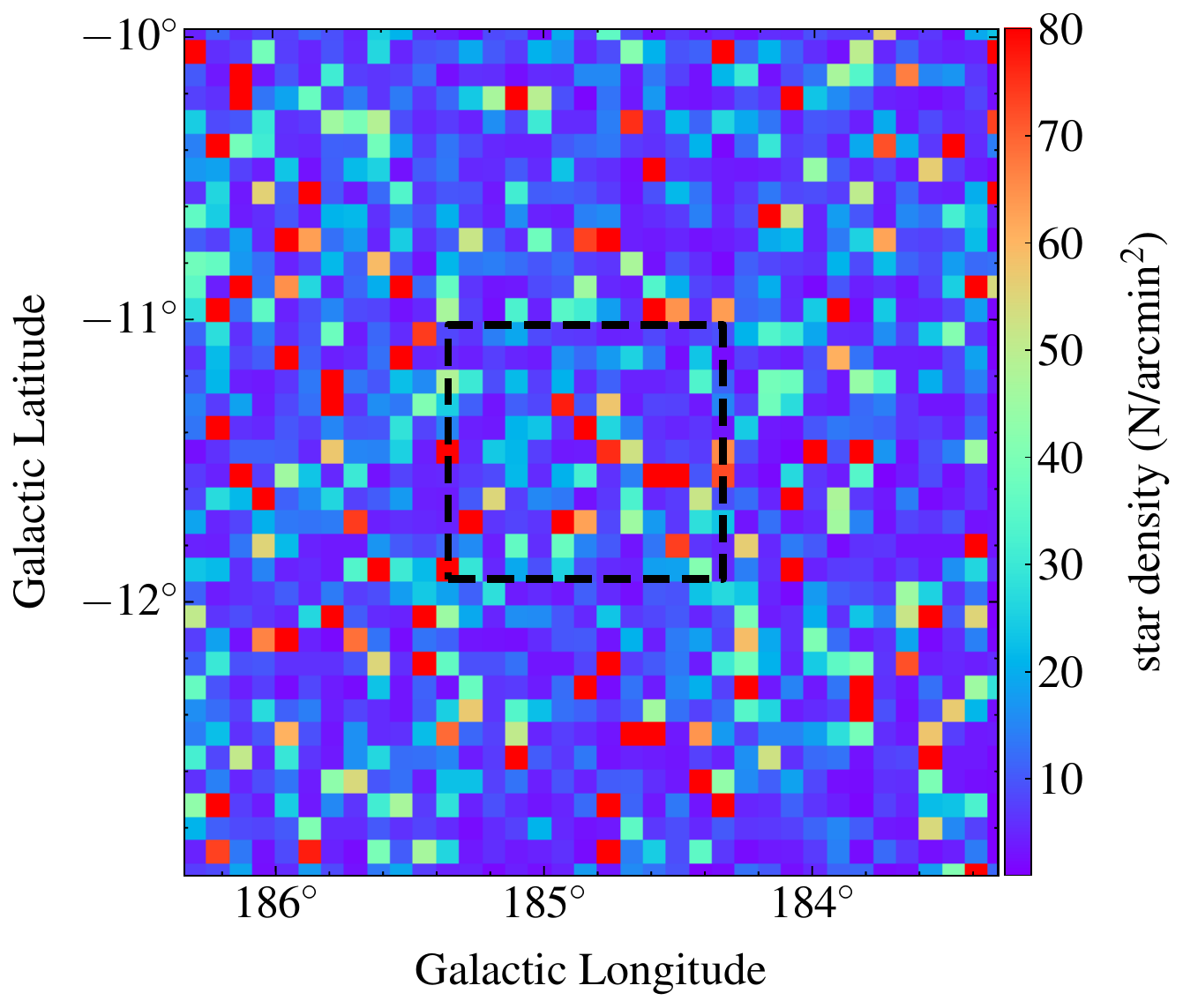}
  \vspace{-3mm}
   \caption{\textbf{Star-density map from  the \textit{Gaia} DR3}. The map was obtained by counting all the stars in the dense region of the clump with a bin of $5^\prime\times5^\prime$. The region enclosed by the black dashed line corresponds to the field of view of Fig.~\ref{contourMaps}c.}
   \label{Gaiaoverdensity}
   \end{figure*}

\begin{figure*}[h!]
   \centering
  \includegraphics[width=9cm, angle=0]{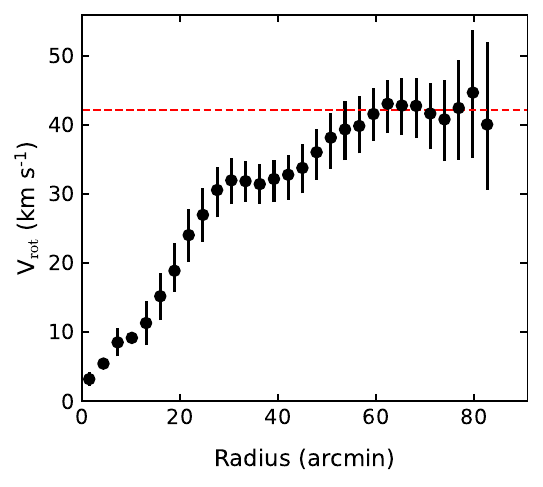}
  \vspace{-3mm}
   \caption{\textbf{Rotational curve profile of the compact clump from the best-fit model.} The error bar of each point is estimated using a Monte Carlo approach. The horizontal red dashed line indicates the rotation velocity estimated by an iterative method from ref.\cite{Lelli+2016}. 
   The process starts by averaging the outermost radial velocities ($V_{\rm N}$ and $V_{\rm N-1}$) to obtain an initial $V_{\rm m}$. Subsequently, it checks if the next inner velocity ($V_{\rm N-2}$) meets the criterion $|V_{\rm N-2} - V_{\rm m}|\leq0.05\times V_{\rm m}$. If so, $V_{\rm N-2}$ is included in recalculating $V_{\rm m}$, and this step is repeated for further inner points. Otherwise, the process terminates, designating the current $V_{\rm m}$ as the rotation velocity.
    }
   \label{RotationCurve}
   \end{figure*}

\begin{figure*}[h!]
   \centering
  \includegraphics[width=11cm, angle=0]{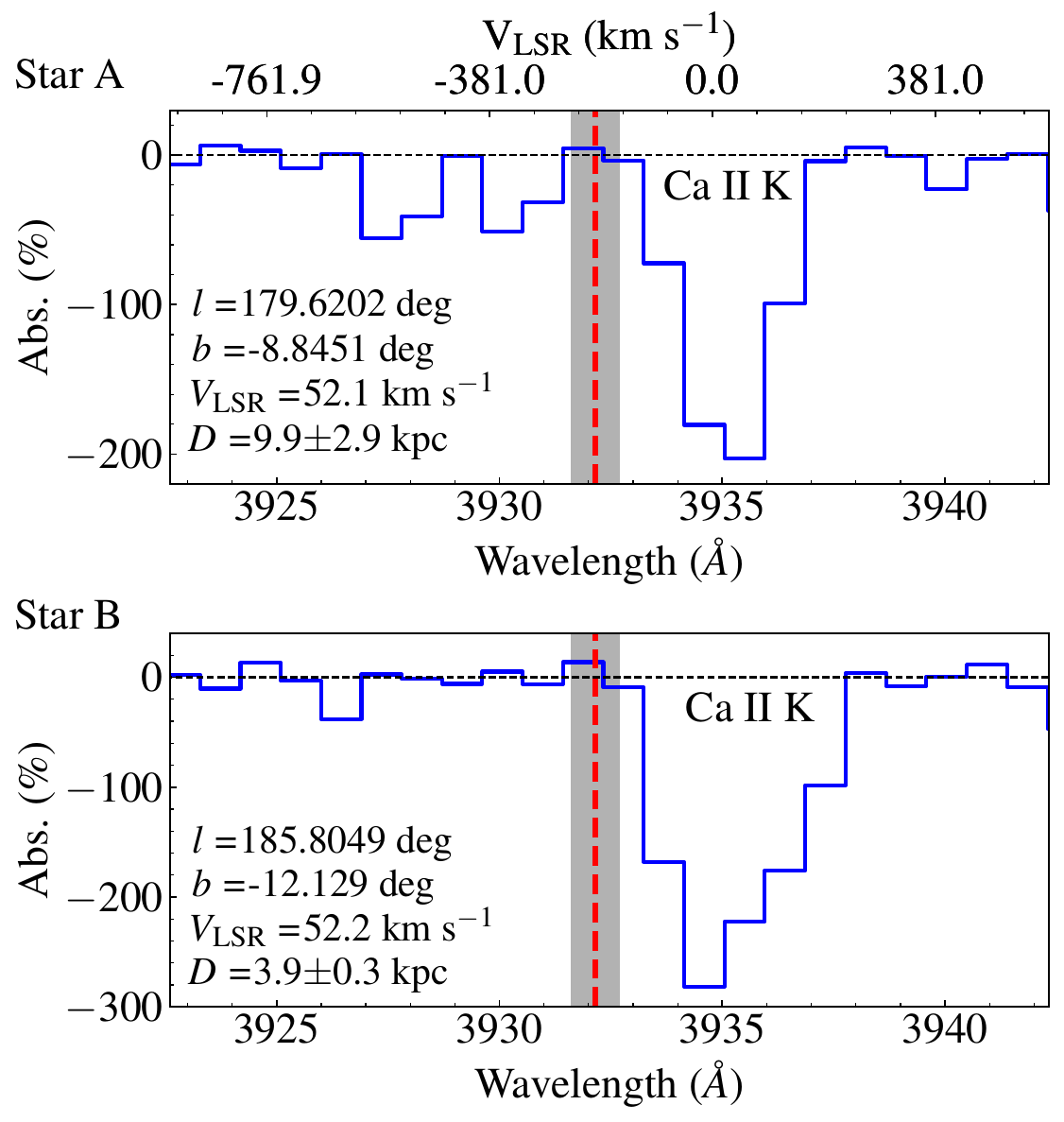}
  \vspace{-3mm}
   \caption{\textbf{The Ca II K absorption line measured towards stars projected on the HVC AC-I.} The optical spectroscopy are obtained from the Large Sky Area Multi-Object Fiber Spectroscopic Telescope (LAMOST) archived data. The equivalent spectral resolution is about 69 \kms. The parameters of each star are listed in the bottom-left corner of the graph. The red dashed line marks the system velocity (-193.0 \kms) of the HVC AC-I, while the gray shaded region indicates its corresponding velocity range. }
   \label{Starabsorbtion}
   \end{figure*}

\clearpage
\begin{table}[h!] 
     \centering
      \caption{\textbf{Obtained and derived properties of the galaxy.} We list: kinematical center with Galactic coordinates ($l\,, b$);  position angle ($PA$); inclination angle ($i$); rotation velocity ($V_{\rm rot}$); velocity dispersion ($V_{\sigma}$); systemic velocity ($V_{\rm LSR}$); line width at $50\%$ level of the peak flux ($W_{50}$); line width at $20\%$ level of the peak flux ($W_{20}$); total \HI\ flux ($S_{\rm t}$); distance derived from the BTF relation ($D$); effective radius of disk ($R_{\rm eff}$); \HI\ gas mass ($M_{\rm H{\sc I}}$); total baryonic mass ($M_{\rm bary}$); total dynamic mass ($M_{\rm dyn}$).}
      \label{tab:Obs}
      \begin{tabular}{lc}
		\\ \hline
Parameter                          & Value                          \\  \hline
$l$                                  & $184.983$                      \\
$b$                                  & $-11.484$                      \\
$PA$ (deg)                         & $34.2_{-1.3}^{+1.3}$          \\
$i$ (deg)                          & $25.9_{-0.6}^{+0.6}$          \\
$V_{\rm rot} (\rm km\,s^{-1})$     & $42.2_{-2.1}^{+2.2}$          \\
$V_{\sigma} (\rm km\,s^{-1})$      & $10.2_{-0.2}^{+0.2}$           \\
$V_{\rm LSR} (\rm km\,s^{-1})$     & $-193.0_{-0.2}^{+0.2}$                \\
$W_{50} (\rm km\,s^{-1})$          & $39.6_{-0.5}^{+0.5}$                  \\
$W_{20} (\rm km\,s^{-1})$          & $60.3_{-0.7}^{+0.7}$                  \\
$S_{\rm t} (\rm 10^3\, Jy\,km\,s^{-1})$  & $6.2_{-0.1}^{+0.1}$               \\ 
$ D \rm(kpc)$                      & $277.7_{-141.6}^{+291.3}$       \\
$R_{\rm eff}$ (kpc)                & $6.7_{-3.4}^{+7.0}$             \\
$M_{\rm H{\sc I}} (\rm 10^8\,M_\odot)$   &$1.1_{-0.8}^{+3.6}$       \\
$M_{\rm bary}(\rm 10^8\,M_\odot)$  & $1.5_{-1.1}^{+4.8}$                  \\
$M_{\rm dyn} (\rm 10^9\,M_\odot)$  & $3.2_{-1.6}^{+3.5}$        \\          
		\hline
	\end{tabular}
\end{table}

\clearpage
\newpage

\section*{Materials and Methods}
\subsection*{FAST HI observations}
New \HI\ (1420.4058 MHz) mapping observations towards the compact gas clump in the HVC AC-I were conducted in March 2024 using the Five-hundred-meter Aperture Spherical radio Telescope (FAST)\cite{Jiang+etal+2019,Jiang+etal+2020}. The mapping observation used the Multi-beam on-the-fly (OTF) mode. This mode is designed to map the sky with 19 beams simultaneously and has a similar scanning trajectory. The observations covered a region of $4^\circ.0\times4^\circ.0$ centering on $lon=185^\circ.056$, $lat =-11^\circ.553$. The 19-beam array has a bandwidth of 500 MHz, spanning from 1000 MHz to 1500 MHz. The half-power beam width (HPBW) of each beam is $2.9^{\prime}$ at 1.4 GHz. The on-sky scanning speed was set to $\rm 15^{\prime\prime} /s$ with an integration time of 1s. Spectra were recorded in the dual-polarization mode of the front end and digitized by the backend Spec(W) with 65536 channels. This corresponds to a frequency resolution of 7.629 kHz at 1.4 GHz and a velocity resolution of 1.611 km s$^{-1}$. During the observations, the system temperature was around 20 K. The pointing accuracy of FAST is better than $8''$. To achieve calibration, a standard noise signal with $T_{\rm cal}\sim$10 K was injected into the signal path at regular intervals of 64s for a period of 1s. The raw data was reduced using the Python-based pipeline HIFAST\cite{Jing+2024}. The final data cube was gridded into pixels of $1^\prime$ by $1^\prime$, achieving a root mean square (RMS) noise of about 2.0 mJy beam$^{-1}$, corresponding to a $3\sigma$ \HI\ column density sensitivity of $\rm 1.2\times10^{18}\,cm^{-2}$ per 20 km s$^{-1}$.

\subsection*{GALFA-HI observations}
\HI\ data for the HVC AC-I were obtained from the Galactic Arecibo L-Band Feed Array \HI\ (GALFA-H\,{\sc i}) survey data release 2 (DR2)\cite{Peek+2018}. This survey utilized the ALFA seven-beam receiver array and covered a total area of 13,000 deg$^2$ spanning $0^{\rm h}<RA<24^{\rm h}$ and $-1^\circ<DEC<38^\circ$, with a spatial resolution of approximately $4^\prime$. Observations encompassed a velocity range from -650 to 650 km s$^{-1}$. The data cube for AC-I, obtained from the wide sets of this survey, exhibits a spectral resolution of 0.754 \kms. It possesses an RMS noise of 0.2 K (21.3 mJy beam$^{-1}$), corresponding to a $3\sigma$ \HI\ column density of $\rm 4.2\times10^{18}\,cm^{-2}$ per 20 km s$^{-1}$.

\subsection*{CO observations}
 To detect the molecular emission in the compact clump, we observed the CO ($J=1-0$) emission at 115.271504 GHz toward the peak \HI\ column density position of the clump using the Purple Mountain Observatory (PMO) 13.7m radio telescope at De Ling Ha in western China, during May and November 2024. The $3\times3$ pixel superconducting spectroscopic array receiver was employed as the front end. A bandwidth of 1 GHz was allocated to the backend, which provides 16,384 channels, resulting in a spectral resolution of 61 kHz. This is equivalent to a velocity resolution of about 0.16 \kms.  The telescope has a half-power beamwidth of approximately $54^{\prime\prime}$ and achieves a pointing accuracy better than $4^{\prime\prime}$. Observations were conducted in single-point mode with a total integration time of approximately 10 hours. The final reduced averaged spectrum was smoothed to a velocity resolution of $1.59\, \rm km\, s^{-1}$, yielding an RMS noise of around 7 mK.

\subsection*{Kinematic analysis of the compact clump}
Since the compact clump in the HVC AC-I shows the characteristics of a rotating disk, we used a 3D tilted ring galaxy-disc model to investigate its kinematics. The fitting disc model is based on 3D-Based Analysis of Rotating Objects via Line Observations ($\rm ^{3D}BAROLO$)\cite{Teodoro+2015}, whose advantage is that it can operate automatically on low-resolution data cubes. Adopting this 3D approach significantly reduces the impact of beam-smearing effects. Utilizing Monte Carlo simulations, the $\rm ^{3D}BAROLO$ model is used to characterize the galaxy structure through a series of concentric rings, which are described by a comprehensive set of geometrical and kinematic parameters. The geometrical parameters encompass the center position ($x_0$, $y_0$) of the galaxy, inclination ($i$), and position angle ($PA$). The kinematic parameters incorporate the systemic velocity ($V_{\rm sys}$), velocity dispersion ($V_\sigma$), and rotation velocity ($V_{\rm rot}$) specifying the circular motion of the rings about the galactic center. 

During the simulations for the rotating disk of the clump, we set the spacing between the rings to be $174^{\prime\prime}$, corresponding to the beam size of FAST. The systemic velocity ($V_{\rm sys}=-193.0$ \kms) and the center position of the clump ($x_0=184^\circ.983, y_0=-11^\circ.484$) were held constant, while other parameters were left free. These values were derived from the integrated \HI\ profile and the \HI\ column density map. The comparison between observations and the final best-fitting model is illustrated through velocity-field (Moment 1) and velocity-position maps (Fig.~\ref{model}). 
There is a missing chunk in the upper-left corner of the galaxy disk. Such a feature may be created by ram-pressure stripping when it falls into the local group. In Fig.~\ref{RotationCurve}, we present the best-fit rotation curve (RC), which exhibits a flat profile in the outer disk. The parameters of the best-fit model are summarized in Tab.~\ref{tab:Obs}, where we determined $PA=34^\circ.2_{-1.3}^{+1.3}$, $i=25^\circ.9_{-0.6}^{+0.6}$, $V_{\rm rot}=42.2_{-2.1}^{+2.2}\, \rm km\, s^{-1}$, $V_\sigma=10.2_{-0.2}^{+0.2}\,\rm km\, s^{-1}$, and the effective radius $R_{\rm eff}=1.4_{-0.1}^{+0.1}$ degrees. The $V_{\rm rot}$ was determined based on the velocity flat part of the RC \cite{Lelli+2016}. The effective radius was corrected for spatial resolution using the equation $R_{\rm eff} = \sqrt{R_{\rm H{\sc I}}^2 - b_{\rm F}^2}$, where $R_{\rm H{\sc I}}$ is the uncorrected radius from the model and $b_{\rm F}$ is the half-beam size of FAST.

\subsection*{Distance and dynamic mass of the compact clump.}
To determine the distance of AC G185.0-11.5, we selected stars from the Gaia point source catalog, constraining them to spectral types O, B, or A to minimize spectral contamination in the stellar Ca II K ion line ($\lambda = 393.478$ nm).  By cross-comparing with a comprehensive star catalog of types O, B, and A \cite{Xiang+2022},  we then chose the most distant star, designated Star A (spectral type B6), located along the tail of HVC AC-I, as shown by the green dot in Fig. \ref{contourMaps}a. The distance to Star A is 9.9$^{+2.9}_{-2.9}$ kpc, derived from parallax measurements. The optical spectrum of Star A was obtained from the Large Sky Area Multi-Object Fiber Spectroscopic Telescope (LAMOST) archive data \cite{Luo+15}. The spectra were re-sampled in constant-velocity pixels, with an equivalent spectral resolution of 69 \kms. The absorption spectrum of the Ca II K line for Star A is shown in Fig. \ref{Starabsorbtion}. No Ca II K absorption ($> 1.5 \sigma$) was detected at the system velocity of HVC AC-I, indicating that the HVC lies behind the star. For comparison, we also identified Star B (spectral type A1), whose spectrum, with a maximum signal-to-noise ratio ($S/N = 24$), is projected onto the compact clump. The distance to Star B is 3.9$^{+0.3}_{-0.3}$ kpc. Similarly, no Ca II K absorption was detected at the system velocity of the HVC, as shown in Fig. \ref{Starabsorbtion}.

For a well-established sample of the gas-rich disk galaxies, a best-fit BTF relation has been obtained as \cite{McGaugh+2015}:
\begin{equation}
    \rm log\,(\it M_{\rm bary})=\rm log\,(\it A)+ s*\rm log\,(\it V_{\rm rot})
\end{equation}
where $\rm log\,(\it A)\rm =1.61\pm0.18$ and $s=4.04\pm0.09$. Using the $^{\rm 3D}$BAROLO model, we determined $V_{\rm rot}$ to be $42.2^{+2.2}_{-2.1}$ km s$^{-1}$ for AC G185.0-11.5. Based on the BTF relation, we calculated the baryonic mass ($M_{\rm bary}$) of the compact clump to be $1.5^{+4.8}_{-1.1}\times10^8\,\rm M_\odot$. The uncertainties in $M_{\rm bary}$ arise primarily from the rotational velocity, incorporating the 2$\sigma$ uncertainties ($\delta y_{\rm all}$) stemming from both the fitting error ($y_{\rm pred}$) and the intrinsic error ($y_{\rm intr}$) of the empirical BTF relation, where $\delta y_{\rm all}=\sqrt{(\delta y_{\rm pred})^{2}+(\delta y_{\rm intr})^{2}}$, which is visually depicted as the gray-shaded area along the BTF line in Fig.~\ref{distance}a. The fitting error $\delta y_{\rm pred}=\sqrt{[\delta \rm log\,(\it A)]^{2}+[\rm log\,(\it V_{\rm rot})\times\delta s]^{2}}$, while the intrinsic error $\delta y_{\rm intr}$ is adopted to be 0.13 from ref.\cite{McGaugh+2015}.

Since no optical counterparts were detected in the clump, we excluded stellar mass from the baryonic mass calculation. Assuming a helium-to-\HI\ ratio (1.33) consistent with Big Bang nucleosynthesis \cite{Planck+20}, the total \HI\ mass was derived as $M_{\rm HI} = 1.1_{-0.8}^{+3.6}\times10^8\,\rm M_{\odot}$. Assuming optically thin \HI\ line emission in the gas clump, we derived its total \HI\ mass using $M_{\rm HI} = 2.36\times10^5 D^2 S_{\rm t}$, where $D$ is the distance in Mpc and $S_{\rm t}$ is the total \HI\ flux (6.2$^{+0.1}_{-0.1}\times10^3\, $Jy) from Gaussian fitting of the integrated \HI\ spectrum (Fig.~\ref{integratedSpectrum}). Consequently, the distance to the clump was determined to be 277.7$^{+291.3}_{-141.6}$ kpc, with uncertainties mainly originating from the baryonic mass, as indicated by the black dashed lines shown in Fig.~\ref{distance}b.

Utilizing the derived values of $V_{\rm rot}$ and $V_\sigma$ from the $\rm ^{3D}BAROLO$ model, we estimated the ratio of $V_{\rm rot}/V_\sigma$ to be  $4.1_{-0.2}^{+0.2}$. This suggests that the compact clump is primarily supported by rotation\cite{Neeleman+etal+2020+nature}. Calculating the dynamical mass, it is also necessary to consider the contribution of velocity dispersion. The dynamical mass can be estimated using the formula $M_{\rm dyn} =({V_{\rm rot}}^2+3{V_{\sigma}}^2)R_{\rm eff}/G$, where $G$ is gravitational constant \cite{Hoffman+1996}. Adopting the distance of 277.7$^{+291.3}_{-141.6}$ kpc, we derived $R_{\rm eff}$ of 6.7$_{-3.4}^{+7.0}$ kpc and $M_{\rm dyn}$ of $3.2_{-1.6}^{+3.5}\times 10^9\,\rm M_\odot$. Finally, we can obtain that $M_{\rm dyn}/M_{\rm bary}$ is 21.6$_{-11.1}^{+22.5}$.

\section*{Acknowledgments}
{This work made use of data from the FAST, a Chinese national mega-science facility, built
and operated by the National Astronomical Observatories, Chinese Academy of Sciences. We thank S. Wang and A. L. Luo for their helpful discussions. 

\subsection*{Funding}
This work is supported by the National Key R\&D Program of China (no 2022YFE0202900 to J.L.X., and C.P.Z.), National Natural Science Foundation of China (nos. 12373001, 12225303, 12421003, 11933011, and  12373011 to J.L.X.,  P.J., Y.X., and X.L.L.),  CAS Project for Young Scientists in Basic Research (no YSBR-063 to P.J.), the Youth Innovation Promotion Association of CAS (to N.P.Y., J.L.X., and P.J.), the Guizhou Provincial Science and Technology Projects (no QKHJC-ZK[2025]MS015 to C.P.Z.), and the Open Project Program of the Key Laboratory of FAST, NAOC, Chinese Academy of Sciences (to L.X.L., J.L.X., P.J., M.Z., C.P.Z., N.P.Y., X.G., and J.J.W.).

\subsection*{Author Contributions}
J.L.X. and X.L.L. led the paper writing. J.L.X. identified the high-velocity clump and led the initial data processing and analysis.  P.J. and M.Z. led the FAST observation.  X.L.L., C.P.Z., and X.G., led the FAST data reduction and analysis. P.J., M.Z., N.P.Y., Y.X., and J.J.W., participated in discussions and the writing of the paper.

\subsection*{Competing Interests}
 The authors declare that they have no competing interests.

\subsection*{Data and Material Availability}
Raw data are available from the FAST data center:  \url{http://fast.bao.ac.cn}. The Gaia, LAMOST and Pan-STARRS1 data  can be downloaded from the archive:   \url{https://gea.esac.esa.int/archive/}, \url{https://www.lamost.org/lmusers/}, and \url{https://outerspace.stsci.edu/display/PANSTARRS/}, respectively. All data needed to evaluate the conclusions of the paper are present in the paper.


\clearpage
\newpage
\begin{thebibliography}{10}
\bibitem{Wakker+1997}
B.~P. {Wakker}, H. {van Woerden}, {High-Velocity Clouds}, {\it \araa\/} {\bf 35}, 217-266 (1997).

\bibitem{Wakker+1991}
B.~P. {Wakker}, H.~{van Woerden}, {Distribution and origin of high-velocity 
clouds. III. Clouds, complexes and populations.}, {\it \aap\/} {\bf 250}, 509-532 (1991).

\bibitem{Putman+2012}
M.~E.~{Putman}, J.~E.~G. {Peek}, M.~R. {Joung}, {Gaseous Galaxy Halos}, {\it \araa\/} {\bf 50}, 491-529 (2012).

\bibitem{Westmeier+2018}
T.~{Westmeier}, {A new all-sky map of Galactic high-velocity clouds from the 21-cm HI4PI survey}, {\it \mnras\/} {\bf 474}, 289-299 (2018).

\bibitem{Richter+2017}
P. {Richter}, S.~E.~{Nuza}, A.~J. {Fox}, B.~P. {Wakker}, N. {Lehner}, N. {Ben Bekhti}, C.~{Fechner}, M.~{Wendt}, J.~C. {Howk}, S. {Muzahid}, R. {Ganguly}, J.~C. {Charlton}, 
{An HST/COS legacy survey of high-velocity ultraviolet absorption in the Milky Way's circumgalactic medium and the Local Group}, {\it \aap\/} {\bf 607}, A48 (2017).

\bibitem{Fox+2019}
A. J. {Fox}, P. {Richter}, T. {Ashley}, T. M. {Heckman}, N.~{Lehner}, J. K.~{Werk}, R.~{Bordoloi}, M. S.~{Peeples}, {The Mass Inflow and Outflow Rates of the Milky Way}, {\it \apj\/} {\bf 884}, 53 (2019).

\bibitem{Woerden+1999+nature}
H.~{van Woerden}, U. J. {Schwarz}, R. F.~{Peletier}, B. P.~{Wakker}, P.~M.~W.~{Kalberla}, {A confirmed location in the Galactic halo for the high-velocity cloud `chain A'}, {\it \nat\/} {\bf 400}, 138-141 (1999).
 
\bibitem{Lucchini+2024}
S.~{Lucchini}, J. J.~{Han}, L. {Hernquist}, C. {Conroy}, {On the Origin of High Velocity Clouds in the Galaxy}, {\it \apj\/} {\bf 974}, 105 (2024).

\bibitem{Mathewson+1966}
D.~S.~{Mathewson}, S.~Y. {Meng}, W.~D. {Brundage}, J.~D. {Kraus}, {Survey of high negative-velocity Hydrogen clouds}, {\it \aj\/} {\bf 71}, 863-864 (1966).

\bibitem{Hulsbosch+1968}
A.~N.~M.~{Hulsbosch}, {High-velocity hydrogen complexes at high galactic latitudes}, {\it \bain\/} {\bf 20}, 33 (1968).

\bibitem{Tamanaha+1997}
C. M.~{Tamanaha}, {The Anticenter Shell and the Anticenter Chain}, {\it \apjss\/} {\bf 109}, 139-175 (1997).

\bibitem{Tamanaha+1996}
C. M.~{Tamanaha}, {Distance Constraints to the Anticenter High-Velocity Clouds}, {\it \apjs\/} {\bf 104}, 81 (1996).

\bibitem{Verschuur+1969}
G.~L.~{Verschuur}, {The High-Velocity Cloud Complexes as Extragalactic Objects in the Local Group}, {\it \apj\/} {\bf 156}, 771 (1969).

\bibitem{Meyerdierks+1991}
H. {Meyerdierks}, {A cloud-Galaxy collision : observation and theory.}, {\it \aap\/} {\bf 251}, 269-275 (1991).

\bibitem{Bruns+2000}
C. {Br{\"u}ns}, J. {Kerp}, P.~M.~W. {Kalberla}, U. {Mebold}, {The head-tail structure of high-velocity clouds. A survey of the northern sky}, {\it \aap\/} {\bf 357}, 120-128 (2000).

\bibitem{Saul+2012}
D. R.~{Saul}, J.~E.~G.~{Peek}, J. {Grcevich}, M.~E. {Putman}, K.~A. {Douglas}, E.~J. {Korpela}, S. {Stanimirovi{\'c}}, C. {Heiles}, S.~J. {Gibson}, M. {Lee}, A. {Begum}, A.~R.~H. {Brown}, B. {Burkhart},  E.~T. {Hamden}, N.~M. {Pingel}, S. {Tonnesen}, {The GALFA-H I Compact Cloud Catalog}, {\it \apj\/} {\bf 758}, 44 (2012).

\bibitem{Ryan-Weber+2008}
E. V.~{Ryan-Weber}, A. {Begum}, T. {Oosterloo}, S. {Pal}, M. J. {Irwin}, V. {Belokurov}, N. W. {Evans}, D. B. {Zucker}, {The Local Group dwarf Leo T: HI on the brink of star formation}, {\it \mnras\/} {\bf 384}, 535-540 (2008).

\bibitem{Giovanelli+2013}
R.~{Giovanelli}, M. P. {Haynes}, E. A.~K. {Adams}, J. M. {Cannon}, K. L.~{Rhode}, J. J. {Salzer}, E. D.~{Skillman}, E. Z.~{Bernstein-Cooper}, K. B.~W.~{McQuinn}, {ALFALFA Discovery of the Nearby Gas-rich Dwarf Galaxy Leo P. I. HI Observations},  {\it \aj\/} {\bf 146}, 15 (2013).

\bibitem{Bruns+2005}
C.~{Br{\"u}ns}, J.~{Kerp}, L.~{Staveley-Smith}, U.~{Mebold}, M.~E.~{Putman}, R.~F.~{Haynes}, P.~M.~W.~{Kalberla}, E.~{Muller}, M.~D.~{Filipovic}, {The Parkes HI Survey of the Magellanic System}, {\it \aap\/} {\bf 432}, 45-67 (2005).

\bibitem{Adams+2013}
E. A.~K.~{Adams}, R.~{Giovanelli}, M. P. {Haynes}, {A Catalog of Ultra-compact High Velocity Clouds from the ALFALFA Survey: Local Group Galaxy Candidates?}, {\it \apj\/} {\bf 768}, 77 (2013).

\bibitem{Blitz+1999}
L. {Blitz}, D. N. {Spergel}, P. J. {Teuben}, D. {Hartmann}, W. B. {Burton}, {High-Velocity Clouds: Building Blocks of the Local Group}, {\it \apj\/} {\bf 514}, 818-843 (1999).

\bibitem{van+1988}
T. {van der Hulst}, R. {Sancisi}, {High-Velocity Gas in M101}, {\it \aj\/} {\bf 95}, 1354 (1988).

\bibitem{Marasco+2012}
A. {Marasco}, F. {Fraternali}, J.~J. {Binney}, {Supernova-driven gas accretion in the Milky Way}, {\it \mnras\/} {\bf 419}, 1107-1120 (2012).

\bibitem{Peek+2007}
J.~E.~G. {Peek}, M.~E. {Putman}, C. F. {McKee}, C. {Heiles}, S. {Stanimirovi{\'c}}, {Reconstructing Deconstruction: High-Velocity Cloud Distance through Disruption Morphology}, {\it \apj\/} {\bf 656}, 907-913 (2007).

\bibitem{Lockman+2008}
F. J. {Lockman}, R. A. {Benjamin}, A.~J. {Heroux}, G. I. {Langston}, {The Smith Cloud: A High-Velocity Cloud Colliding with the Milky Way}, {\it \apjl\/} {\bf 679}, L21 (2008).

\bibitem{Genzel+etal+2006+nature}
R.~{Genzel}, L.~J.~{Tacconi}, F. {Eisenhauer}, N.~M.~{F{\"o}rster Schreiber}, A. {Cimatti}, E. {Daddi}, N. {Bouch{\'e}}, R. {Davies}, M.~D. {Lehnert}, D. {Lutz}, N. {Nesvadba}, A. {Verma}, R. {Abuter}, K.~{Shapiro}, A. {Sternberg}, A. {Renzini}, X. {Kong}, N. {Arimoto}, M. {Mignoli}, {The rapid formation of a large rotating disk galaxy three billion years after the Big Bang}, {\it \nat\/} {\bf 442}, 786-789 (2006).

\bibitem{Lu+etal+2022+NA}
X. {Lu}, G. X. {Li}, Q. Z. {Zhang}, Y. X. {Lin}, {A massive Keplerian protostellar disk with flyby-induced spirals in the Central Molecular Zone}, {\it Nature Astronomy\/} {\bf 6}, 837-843 (2022).

\bibitem{Neeleman+etal+2020+nature}
M.~{Neeleman}, J. X. {Prochaska}, N. {Kanekar}, M. {Rafelski}, {A cold, massive, rotating disk galaxy 1.5 billion years after the Big Bang}, {\it \nat\/} {\bf 581}, 269-272 (2020).

\bibitem{Rizzo+etal+2020+nature}
F.~{Rizzo}, S.~{Vegetti}, D.~{Powell}, F.~{Fraternali}, J.~P.~{McKean}, H.~R. {Stacey}, S.~D.~M. {White}, {A dynamically cold disk galaxy in the early Universe}, {\it \nat\/} {\bf 584}, 201-204 (2020).

\bibitem{Thom+2008}
C.~{Thom}, J. E. G.~{Peek}, M. E.~{Putman}, C.~{Heiles}, K. M.~G.~{Peek}, R.~{Wilhelm}, {An Accurate Distance to High-Velocity Cloud Complex C}, {\it \apj\/} {\bf 684}, 364-372 (2008).

\bibitem{Lehner+2011+sci}
N. {Lehner}, J. C. {Howk}, {A Reservoir of Ionized Gas in the Galactic Halo to Sustain Star Formation in the Milky Way}, {\it Science\/} {\bf 334}, 955 (2011).

\bibitem{Lehner+2022}
N. {Lehner}, J. C. {Howk},  A. {Marasco}, F. {Fraternali}, {Intermediate- and high-velocity clouds in the Milky Way-I. Covering factors and vertical heights}, {\it \mnras\/} {\bf 513}, 3228-3240 (2022).

\bibitem{Collaboration+2023}
{{Gaia Collaboration}, A. {Vallenari}, A.~G.~A. {Brown}, T. {Prusti}, J.~H.~J. {de Bruijne}, F. {Arenou}, C. {Babusiaux}, M. {Biermann}, O.~L. {Creevey}, C. {Ducourant}, D.~W. {Evans}, L. {Eyer}, R. {Guerra}, A. {Hutton}, C. {Jordi}, S.~A. {Klioner}, U.~L. {Lammers}, L. {Lindegren}, X. {Luri}, F. {Mignard}, C. {Panem}, D. {Pourbaix}, S. {Randich}, P. {Sartoretti}, C. {Soubiran}, P. {Tanga}, N.~A. {Walton}, C.~A.~L. {Bailer-Jones}, U. {Bastian}, R. {Drimmel}, F. {Jansen}, D. {Katz}, M.~G. {Lattanzi}, F. {van Leeuwen}, J. {Bakker}, C. {Cacciari}, J. {Casta{\~n}eda}, F. {De Angeli}, C. {Fabricius}, M. {Fouesneau}, Y. {Fr{\'e}mat}, L. {Galluccio}, A. {Guerrier}, U. {Heiter}, E. {Masana}, R. {Messineo}, N. {Mowlavi}, C. {Nicolas}, K. {Nienartowicz}, F. {Pailler}, P. {Panuzzo}, F. {Riclet}, W. {Roux}, G.~M. {Seabroke}, R. {Sordo}, F. {Th{\'e}venin}, G. {Gracia-Abril}, J. {Portell}, D. {Teyssier}, M. {Altmann}, R. {Andrae}, M. {Audard}, I. {Bellas-Velidis}, K. {Benson}, J. {Berthier}, R. {Blomme}, P.~W. {Burgess}, D. {Busonero}, G. {Busso}, H. {C{\'a}novas}, B. {Carry}, A. {Cellino}, N. {Cheek}, G. {Clementini}, Y. {Damerdji}, M. {Davidson}, P. {de Teodoro}, M. {Nu{\~n}ez Campos}, L. {Delchambre}, A. {Dell'Oro}, P. {Esquej}, J. {Fern{\'a}ndez-Hern{\'a}ndez}, E. {Fraile}, D. {Garabato}, P. {Garc{\'\i}a-Lario}, E. {Gosset}, R. {Haigron}, J. -L. {Halbwachs}, N.~C. {Hambly}, D.~L. {Harrison}, J. {Hern{\'a}ndez}, D. {Hestroffer}, S.~T. {Hodgkin}, B. {Holl}, K. {Jan{\ss}en}, G. {Jevardat de Fombelle}, S. {Jordan}, A. {Krone-Martins}, A.~C. {Lanzafame}, W. {L{\"o}ffler}, O. {Marchal}, P.~M. {Marrese}, A. {Moitinho}, K. {Muinonen}, P. {Osborne}, E. {Pancino}, T. {Pauwels}, A. {Recio-Blanco}, C. {Reyl{\'e}}, M. {Riello}, L. {Rimoldini}, T. {Roegiers}, J. {Rybizki}, L.~M. {Sarro}, C. {Siopis}, M. {Smith}, A. {Sozzetti}, E. {Utrilla}, M. {van Leeuwen}, U. {Abbas}, P. {{\'A}brah{\'a}m}, A. {Abreu Aramburu}, C. {Aerts}, J.~J. {Aguado}, M. {Ajaj}, F. {Aldea-Montero}, G. {Altavilla}, M.~A. {{\'A}lvarez}, J. {Alves}, F. {Anders}, R.~I. {Anderson}, E. {Anglada Varela}, T. {Antoja}, D. {Baines}, S.~G. {Baker}, L. {Balaguer-N{\'u}{\~n}ez}, E. {Balbinot}, Z. {Balog}, C. {Barache}, D. {Barbato}, M. {Barros}, M.~A. {Barstow}, S. {Bartolom{\'e}}, J. L. {Bassilana}, N. {Bauchet}, U. {Becciani}, M. {Bellazzini}, A. {Berihuete}, M. {Bernet}, S. {Bertone}, L. {Bianchi}, A. {Binnenfeld}, S. {Blanco-Cuaresma}, A. {Blazere}, T. {Boch}, A. {Bombrun}, D. {Bossini}, S. {Bouquillon}, A. {Bragaglia}, L. {Bramante}, E. {Breedt}, A. {Bressan}, N. {Brouillet}, E. {Brugaletta}, B. {Bucciarelli}, A. {Burlacu}, A.~G. {Butkevich}, R. {Buzzi}, E. {Caffau}, R. {Cancelliere}, T. {Cantat-Gaudin}, R. {Carballo}, T. {Carlucci}, M.~I. {Carnerero}, J.~M. {Carrasco}, L. {Casamiquela}, M. {Castellani}, A. {Castro-Ginard}, L. {Chaoul}, P. {Charlot}, L. {Chemin}, V. {Chiaramida}, A. {Chiavassa}, N. {Chornay}, G. {Comoretto}, G. {Contursi}, W.~J. {Cooper}, T. {Cornez}, S. {Cowell}, F. {Crifo}, M. {Cropper}, M. {Crosta}, C. {Crowley}, C. {Dafonte}, A. {Dapergolas}, M. {David}, P. {David}, P. {de Laverny}, F. {De Luise}, R. {De March}, J. {De Ridder}, R. {de Souza}, A. {de Torres}, E.~F. {del Peloso}, E. {del Pozo}, M. {Delbo}, A. {Delgado}, J. -B. {Delisle}, C. {Demouchy}, T.~E. {Dharmawardena}, P. {Di Matteo}, S. {Diakite}, C. {Diener}, E. {Distefano}, C. {Dolding}, B. {Edvardsson}, H. {Enke}, C. {Fabre}, M. {Fabrizio}, S. {Faigler}, G. {Fedorets}, P. {Fernique}, A. {Fienga}, F. {Figueras}, Y. {Fournier}, C. {Fouron}, F. {Fragkoudi}, M. {Gai}, A. {Garcia-Gutierrez}, M. {Garcia-Reinaldos}, M. {Garc{\'\i}a-Torres}, A. {Garofalo}, A. {Gavel}, P. {Gavras}, E. {Gerlach}, R. {Geyer}, P. {Giacobbe}, G. {Gilmore}, S. {Girona}, G. {Giuffrida}, R. {Gomel}, A. {Gomez}, J. {Gonz{\'a}lez-N{\'u}{\~n}ez}, I. {Gonz{\'a}lez-Santamar{\'\i}a}, J.~J. {Gonz{\'a}lez-Vidal}, M. {Granvik}, P. {Guillout}, J. {Guiraud}, R. {Guti{\'e}rrez-S{\'a}nchez}, L.~P. {Guy}, D. {Hatzidimitriou}, M. {Hauser}, M. {Haywood}, A. {Helmer}, A. {Helmi}, M.~H. {Sarmiento}, S.~L. {Hidalgo}, T. {Hilger}, N. {H{\l}adczuk}, D. {Hobbs}, G. {Holland}, H.~E. {Huckle}, K. {Jardine}, G. {Jasniewicz}, A. {Jean-Antoine Piccolo}, {\'O}. {Jim{\'e}nez-Arranz}, A. {Jorissen}, J. {Juaristi Campillo}, F. {Julbe}, L. {Karbevska}, P. {Kervella}, S. {Khanna}, M. {Kontizas}, G. {Kordopatis}, A.~J. {Korn}, {\'A}. {K{\'o}sp{\'a}l}, Z. {Kostrzewa-Rutkowska}, K. {Kruszy{\'n}ska}, M. {Kun}, P. {Laizeau}, S. {Lambert}, A.~F. {Lanza}, Y. {Lasne}, J. -F. {Le Campion}, Y. {Lebreton}, T. {Lebzelter}, S. {Leccia}, N. {Leclerc}, I. {Lecoeur-Taibi}, S. {Liao}, E.~L. {Licata}, H.~E.~P. {Lindstr{\o}m}, T.~A. {Lister}, E. {Livanou}, A. {Lobel}, A. {Lorca}, C. {Loup}, P. {Madrero Pardo}, A. {Magdaleno Romeo}, S. {Managau}, R.~G. {Mann}, M. {Manteiga}, J.~M. {Marchant}, M. {Marconi}, J. {Marcos}, M.~M.~S. {Marcos Santos}, D. {Mar{\'\i}n Pina}, S. {Marinoni}, F. {Marocco}, D.~J. {Marshall}, L. {Martin Polo}, J.~M. {Mart{\'\i}n-Fleitas}, G. {Marton}, N. {Mary}, A. {Masip}, D. {Massari}, A. {Mastrobuono-Battisti}, T. {Mazeh}, P.~J. {McMillan}, S. {Messina}, D. {Michalik}, N.~R. {Millar}, A. {Mints}, D. {Molina}, R. {Molinaro}, L. {Moln{\'a}r}, G. {Monari}, M. {Mongui{\'o}}, P. {Montegriffo}, A. {Montero}, R. {Mor}, A. {Mora}, R. {Morbidelli}, T. {Morel}, D. {Morris}, T. {Muraveva}, C.~P. {Murphy}, I. {Musella}, Z. {Nagy}, L. {Noval}, F. {Oca{\~n}a}, A. {Ogden}, C. {Ordenovic}, J.~O. {Osinde}, C. {Pagani}, I. {Pagano}, L. {Palaversa}, P.~A. {Palicio}, L. {Pallas-Quintela}, A. {Panahi}, S. {Payne-Wardenaar}, X. {Pe{\~n}alosa Esteller}, A. {Penttil{\"a}}, B. {Pichon}, A.~M. {Piersimoni}, F. X. {Pineau}, E. {Plachy}, G. {Plum}, E. {Poggio}, A. {Pr{\v{s}}a}, L. {Pulone}, E. {Racero}, S. {Ragaini}, M. {Rainer}, C.~M. {Raiteri}, N. {Rambaux}, P. {Ramos}, M. {Ramos-Lerate}, P. {Re Fiorentin}, S. {Regibo}, P.~J. {Richards}, C. {Rios Diaz}, V. {Ripepi}, A. {Riva}, H. -W. {Rix}, G. {Rixon}, N. {Robichon}, A.~C. {Robin}, C. {Robin}, M. {Roelens}, H.~R.~O. {Rogues}, L. {Rohrbasser}, M. {Romero-G{\'o}mez}, N. {Rowell}, F. {Royer}, D. {Ruz Mieres}, K.~A. {Rybicki}, G. {Sadowski}, A. {S{\'a}ez N{\'u}{\~n}ez}, A. {Sagrist{\`a} Sell{\'e}s}, J. {Sahlmann}, E. {Salguero}, N. {Samaras}, V. {Sanchez Gimenez}, N. {Sanna}, R. {Santove{\~n}a}, M. {Sarasso}, M. {Schultheis}, E. {Sciacca}, M. {Segol}, J.~C. {Segovia}, D. {S{\'e}gransan}, D. {Semeux}, S. {Shahaf}, H.~I. {Siddiqui}, A. {Siebert}, L. {Siltala}, A. {Silvelo}, E. {Slezak}, I. {Slezak}, R.~L. {Smart}, O.~N. {Snaith}, E. {Solano}, F. {Solitro}, D. {Souami}, J. {Souchay}, A. {Spagna}, L. {Spina}, F. {Spoto}, I.~A. {Steele}, H. {Steidelm{\"u}ller}, C.~A. {Stephenson}, M. {S{\"u}veges}, J. {Surdej}, L. {Szabados}, E. {Szegedi-Elek}, F. {Taris}, M.~B. {Taylor}, R. {Teixeira}, L. {Tolomei}, N. {Tonello}, F. {Torra}, J. {Torra}, G. {Torralba Elipe}, M. {Trabucchi}, A.~T. {Tsounis}, C. {Turon}, A. {Ulla}, N. {Unger}, M.~V. {Vaillant}, E. {van Dillen}, W. {van Reeven}, O. {Vanel}, A. {Vecchiato}, Y. {Viala}, D. {Vicente}, S. {Voutsinas}, M. {Weiler}, T. {Wevers}, {\L}. {Wyrzykowski}, A. {Yoldas}, P. {Yvard}, H. {Zhao}, J. {Zorec}, S. {Zucker}, T. {Zwitter}, {Gaia Data Release 3. Summary of the content and survey properties}, {\it \aap\/} {\bf 674}, A1 (2023).
}

\bibitem{Tully-Fisher+1977}
R.~B. {Tully}, J.~R. {Fisher}, {A new method of determining distances to galaxies.}, {\it \aap\/} {\bf 54}, 661-673 (1977).

\bibitem{Makarov+2014}
D.~{Makarov}, P. {Prugniel}, N. {Terekhova}, H. {Courtois}, I. {Vauglin}, {HyperLEDA. III. The catalogue of extragalactic distances}, {\it \aap\/} {\bf 570}, A13 (2014).

\bibitem{McGaugh+2000}
S.~S.~{McGaugh}, J.~M. {Schombert}, G.~D. {Bothun}, W.~J.~G. {de Blok}, {The Baryonic Tully-Fisher Relation}, {\it \apjl\/} {\bf 533}, L99-L102 (2000).

\bibitem{McGaugh+2012}
S.~S.~{McGaugh}, {The Baryonic Tully-Fisher relation of gas-rich galaxies as a Test of $\Lambda$CDM and MOND}, {\it \aj\/} {\bf 143}, 40 (2012).

\bibitem{McGaugh+2015}
S. S.~{McGaugh}, J. M.~{Schombert}, {Weighing Galaxy Disks With the Baryonic Tully-Fisher Relation}, {\it \apj\/} {\bf 802}, 18 (2015).

\bibitem{Lelli+2016}
F.~{Lelli}, S. S.~{McGaugh}, J. M. {Schombert}, {The Small Scatter of the Baryonic Tully-Fisher Relation}, {\it \apjl\/} {\bf 816}, L14 (2016).

\bibitem{Nicastro+2003+NA}
F. {Nicastro}, A. {Zezas}, M. {Elvis}, S. {Mathur}, F. {Fiore}, C. {Cecchi-Pestellini}, D. {Burke}, J. {Drake}, P. {Casella}, {The far-ultraviolet signature of the `missing' baryons in the Local Group of galaxies}, {\it \nat\/} {\bf 421}, 719-721 (2003).

\bibitem{Davies+2006}
J.~I.~{Davies}, M.~J.~{Disney}, R.~F. {Minchin}, R.~{Auld}, R. {Smith}, {The existence and detection of optically dark galaxies by 21-cm surveys}, {\it \mnras\/} {\bf 368}, 1479-1488 (2006).

\bibitem{Magnani-Onello+1995}
L. {Magnani}, J.~S. {Onello}, {A New Method for Determining the CO to H$_{2}$ Conversion Factor for Translucent Clouds}, {\it \apj\/} {\bf 443}, 169 (1995).

\bibitem{Chambers+2016}
K.~C.~{Chambers}, E.~A. {Magnier}, N.~{Metcalfe}, H.~A. {Flewelling}, M.~E. {Huber}, C.~Z. {Waters}, L. {Denneau}, P.~W. {Draper}, D. {Farrow}, D.~P. {Finkbeiner}, C. {Holmberg}, J. {Koppenhoefer},  P.~A.~{Price}, A. {Rest}, R.~P. {Saglia}, E.~F. {Schlafly}, S.~J. {Smartt}, W. {Sweeney}, R.~J. {Wainscoat}, W.~S. {Burgett}, S. {Chastel}, T. {Grav}, J.~N. {Heasley}, K.~W. {Hodapp}, R. {Jedicke}, N.~{Kaiser}, R. P. {Kudritzki}, G.~A.~{Luppino}, R.~H.~{Lupton}, D.~G.~{Monet}, J.~S.~{Morgan}, P.~M.~{Onaka}, B. {Shiao}, C.~W. {Stubbs}, J.~L. {Tonry}, R. {White}, E. {Ba{\~n}ados}, E.~F. {Bell},  R.~{Bender}, E.~J. {Bernard}, M. {Boegner}, F. {Boffi}, M.~T. {Botticella}, A. {Calamida}, S. {Casertano}, W. P. {Chen}, X. {Chen}, S. {Cole}, N. {Deacon}, C. {Frenk}, A. {Fitzsimmons}, S. {Gezari},  V.~{Gibbs}, C. {Goessl}, T. {Goggia}, R. {Gourgue}, B. {Goldman}, P. {Grant}, E.~K. {Grebel}, N.~C. {Hambly}, G. {Hasinger}, A.~F. {Heavens}, T.~M. {Heckman}, R. {Henderson}, T. {Henning}, M. {Holman},  U.~{Hopp}, W. -H.~{Ip}, S. {Isani}, M. {Jackson}, C.~D. {Keyes}, A.~M. {Koekemoer}, R. {Kotak}, D. {Le}, D. {Liska}, K.~S. {Long}, J.~R. {Lucey}, M. {Liu}, N.~F. {Martin}, G. {Masci}, B. {McLean},  E.~{Mindel}, P. {Misra}, E. {Morganson}, D.~N.~A. {Murphy}, A. {Obaika}, G. {Narayan}, M.~A. {Nieto-Santisteban}, P. {Norberg}, J.~A. {Peacock}, E.~A. {Pier}, M. {Postman}, N. {Primak}, C. {Rae}, A. {Rai},  A.~{Riess}, A. {Riffeser}, H.~W.~{Rix}, S.~{R{\"o}ser}, R. {Russel}, L. {Rutz}, E. {Schilbach}, A.~S.~B. {Schultz}, D. {Scolnic}, L. {Strolger}, A. {Szalay}, S. {Seitz}, E. {Small}, K.~W. {Smith},  D.~R.~{Soderblom}, P.~{Taylor}, R. {Thomson}, A.~N. {Taylor}, A.~R. {Thakar}, J. {Thiel}, D. {Thilker}, D. {Unger}, Y. {Urata}, J. {Valenti}, J. {Wagner}, T. {Walder}, F. {Walter}, S.~P. {Watters},  S.~{Werner}, W.~M. {Wood-Vasey}, R. {Wyse}, {The Pan-STARRS1 Surveys}, arXiv:1612.05560 (2016).

\bibitem{Ponomareva+2017}
A. A.~{Ponomareva}, M. A.~W.~{Verheijen}, R. F.~{Peletier}, A. {Bosma}, {The multiwavelength Tully-Fisher relation with spatially resolved HI kinematics}, {\it \mnras\/} {\bf 469}, 2387-2400 (2017).

\bibitem{Green+19}
G. M.~{Green}, E.~{Schlafly}, C. {Zucker}, J. S. {Speagle}, D. {Finkbeiner}, {A 3D Dust Map Based on Gaia, Pan-STARRS 1, and 2MASS}, {\it \apj\/} {\bf 887}, 93 (2019).

\bibitem{Duc+04}
P. A.~{Duc}, F.~{Bournaud}, F. {Masset}, {A top-down scenario for the formation of massive Tidal Dwarf Galaxies}, {\it \aap\/} {\bf 427}, 803-814 (2004).

\bibitem{Lelli+15}
F. {Lelli}, P.~A. {Duc}, E.~{Brinks}, F. {Bournaud}, S. S. {McGaugh}, U. {Lisenfeld}, P. M. {Weilbacher}, M. {Boquien}, Y. {Revaz}, J.~{Braine}, B. S.~{Koribalski}, P. E. {Belles}, {Gas dynamics in tidal dwarf galaxies: Disc formation at z = 0}, {\it \aap\/} {\bf 584}, A113 (2015).

\bibitem{van+22}
P.~{van Dokkum}, Z. {Shen}, M. A.~{Keim}, S.~{Trujillo-Gomez}, S.~{Danieli}, D.~{Dutta Chowdhury}, R. {Abraham}, C.~{Conroy}, J.~M. D.~{Kruijssen}, D. {Nagai}, A.~{Romanowsky}, {A trail of dark-matter-free galaxies from a bullet-dwarf collision}, {\it \nat\/} {\bf 605}, 435-439 (2022).

\bibitem{Llambay+17}
A.~{Ben{\'\i}tez-Llambay}, J. F.~{Navarro}, C. S.~{Frenk}, T.~{Sawala}, K.~{Oman}, A. {Fattahi}, M.~{Schaller}, J.~{Schaye}, R. A. {Crain}, T. {Theuns}, {The properties of `dark' $\Lambda$CDM haloes in the Local Group}, {\it \mnras\/} {\bf 465}, 3913-3926 (2017).

\bibitem{Roman+21}
J.~{Rom{\'a}n}, M. G. {Jones}, M.~{Montes}, L.~{Verdes-Montenegro}, J. {Garrido}, S.~{S{\'a}nchez}, {A diffuse tidal dwarf galaxy destined to fade out as a ``dark galaxy"}, {\it \aap\/} {\bf 649}, L14 (2021).

\bibitem{Simon+2006}
J. D. {Simon}, L. {Blitz}, A. A. {Cole}, M. D. {Weinberg}, M. {Cohen}, {The Cosmological Significance of High-Velocity Cloud Complex H}, {\apj\/} {\bf 640}, 270-281 (2006).

\bibitem{Robishaw+2002}
T. {Robishaw}, J. D. {Simon}, L. {Blitz}, {H I Imaging of LGS 3 and an Apparently Interacting High-Velocity Cloud}, {\apjl\/} {\bf 580}, L129-L132 (2002).

\bibitem{McConnachie+2012}
A. W. {McConnachie}, {The Observed Properties of Dwarf Galaxies in and around the Local Group}, {\aj\/} {\bf 144}, 4 (2012).

\bibitem{Klypin+2002}
A. {Klypin}, H. S. {Zhao}, R. S. {Somerville}, {{\ensuremath{\Lambda}}CDM-based Models for the Milky Way and M31. I. Dynamical Models}, {\apj\/} {\bf 573}, 597-613 (2002).

\bibitem{Roediger+2008}
E. {Roediger}, M. {Br{\"u}ggen}, {Ram-pressure stripping of disc galaxies orbiting in clusters - II. Galactic wakes}, {\mnras\/} {\bf 388}, 465-486 (2008).

\bibitem{Plockinger+2012}
S. {Pl{\"o}ckinger}, G. {Hensler}, {Do high-velocity clouds trace the dark matter subhalo population?}, {\aap\/} {\bf 547}, A43 (2012).

\bibitem{Klypin+99}
A.~{Klypin}, A. V.~{Kravtsov}, O. {Valenzuela}, F. {Prada}, {Where Are the Missing Galactic Satellites?}, {\it \apj\/} {\bf 522}, 82-92 (1999).

\bibitem{Moore+99}
B.~{Moore}, S.~{Ghigna}, F.~{Governato}, G.~{Lake}, T.~{Quinn}, J.~{Stadel}, P.~{Tozzi}, {Dark Matter Substructure within Galactic Halos}, {\it \apjl\/} {\bf 524}, L19-L22 (1999).

\bibitem{Jiang+etal+2019}
P.~{Jiang}, Y. L. {Yue}, H. Q. {Gan}, R. {Yao}, H. {Li}, G. F. {Pan}, J. H. {Sun}, D. J. {Yu}, H. F. {Liu}, N. Y. {Tang}, L. {Qian}, J. G. {Lu}, J. {Yan}, B. {Peng}, S. X. {Zhang}, Q. M. {Wang}, Q. {Li},   D.~{Li}, FAST Collaboration, {Commissioning progress of the FAST}, {\it Science China Physics, Mechanics, and Astronomy\/} {\bf 62}, 959502 (2019).
       
\bibitem{Jiang+etal+2020}
P.~{Jiang}, N. Y.~{Tang}, L. G. {Hou}, M. T. {Liu}, M. {Kr{\v{c}}o}, L. {Qian}, J. H. {Sun}, T. C. {Ching}, B. {Liu}, Y. {Duan}, Y. L. {Yue}, H. Q. {Gan}, R. {Yao}, H. {Li}, G. F. {Pan}, D. J. {Yu}, H. F. {Liu}, D. {Li}, B. {Peng}, J.~{Yan}, {FAST Collaboration}, {The fundamental performance of FAST with 19-beam receiver at L band}, {\it Research in Astronomy and Astrophysics\/} {\bf 20}, 064 (2020).
    
\bibitem{Jing+2024}
Y. J.~{Jing}, J.~{Wang}, C. {Xu}, Z. M. {Liu}, Q. Z.~{Chen}, T. T. {Liang}, J. L. {Xu}, Y. X.~{Cao}, J.~{Wang}, H. J.~{Hu}, C. P.~{Zhang}, Q. {Guo}, L.~{Gao}, M. {Ai}, H. Q. {Gan}, X. Y. {Gao}, J. L. {Han}, L. G.~{Hou}, Z. P. {Hou}, P. {Jiang}, X. {Kong}, F. J. {Li}, Z. R. {Liu}, L. {Shao}, H. X. {Pan}, J. {Pan}, L. {Qian}, J. H. {Sun}, N. Y.~{Tang}, Q. L. {Yang}, B.~{Zhang}, Z. Y. {Zhang}, M.~{Zhu}, {HiFAST: An HI data calibration and imaging pipeline for FAST}, {\it Science China Physics, Mechanics, and Astronomy\/} {\bf 67}, 259514 (2024).
     
\bibitem{Peek+2018}
J.~E.~G.~{Peek}, B. L. {Babler}, Y. {Zheng}, S.~E. {Clark}, K. A. {Douglas}, E. J. {Korpela}, M.~E. {Putman}, S.~{Stanimirovi{\'c}}, S. J. {Gibson}, C. {Heiles}, {The GALFA-HI Survey Data Release 2},  {\it \apjss\/} {\bf 234}, 2 (2018).
                 
\bibitem{Teodoro+2015}
E.~M.~{Di Teodoro}, F. {Fraternali}, {$^{\rm 3D}$BAROLO: a new 3D algorithm to derive rotation curves of galaxies}, {\it \mnras\/} {\bf 451}, 3021-3033 (2015).

\bibitem{Xiang+2022}
S. M.~{Xiang}, H. W. {Rix}, Y. S. {Ting}, R.~P. {Kudritzki}, C. {Conroy}, E. {Zari}, J.~R. {Shi}, N.~{Przybilla}, M. {Ramirez-Tannus}, A. {Tkachenko}, S.~{Gebruers}, X. W. {Liu}, {Stellar labels for hot stars from low-resolution spectra. I. The HotPayne method and results for 330 000 stars from LAMOST DR6}, {\it \aap\/} {\bf 662}, A66 (2022).

\bibitem{Luo+15}
A. L. {Luo}, Y. H. {Zhao}, G. {Zhao}, L. C. {Deng}, X. W. {Liu}, Y. P. {Jing}, G. {Wang}, H. T. {Zhang}, J. R. {Shi}, X. Q. {Cui}, Y. Q. {Chu}, G. P. {Li}, Z. R. {Bai}, Y. {Wu}, Y. {Cai}, S. Y. {Cao}, Z. H. {Cao}, J. L. {Carlin}, H. Y. {Chen}, J. J. {Chen}, K. X. {Chen}, L. {Chen}, X. L. {Chen}, X. Y. {Chen}, Y. {Chen}, N. {Christlieb}, J. R. {Chu}, C. Z. {Cui}, Y. Q. {Dong}, B. {Du}, D. W. {Fan}, L. {Feng}, J. N. {Fu}, P. {Gao}, X. F. {Gong}, B. Z. {Gu}, Y. X. {Guo}, Z. W. {Han}, B. L. {He}, J. L. {Hou}, Y. H. {Hou}, W. {Hou}, H. Z. {Hu}, N. S. {Hu}, Z. W. {Hu}, Z. Y. {Huo}, L. {Jia}, F. H. {Jiang}, X. {Jiang}, Z. B. {Jiang}, G. {Jin}, X. {Kong}, X. {Kong}, Y. J. {Lei}, A. H. {Li}, C. H. {Li}, G. W. {Li}, H. N. {Li}, J. {Li}, Q. {Li}, S. {Li}, S. S. {Li}, X. N. {Li}, Y. {Li}, Y. B. {Li}, Y. P. {Li}, Y. {Liang}, C. C. {Lin}, C. {Liu}, G. R. {Liu}, G. Q. {Liu}, Z. G. {Liu}, W. Z. {Lu}, Y. {Luo}, Y. D. {Mao}, H. {Newberg}, J. J. {Ni}, Z. X. {Qi}, Y. J. {Qi}, S. Y. {Shen}, H. M. {Shi}, J. {Song}, Y. H. {Song}, D. Q. {Su}, H. J. {Su}, Z. H. {Tang}, Q. S. {Tao}, Y. {Tian}, D. {Wang}, D. Q. {Wang}, F. F. {Wang}, G. M. {Wang}, H. {Wang}, H. C. {Wang}, J. {Wang}, J. N. {Wang}, J. L. {Wang}, J. P. {Wang}, J. X. {Wang}, L. {Wang}, M. X. {Wang}, S. G. {Wang}, S. Q. {Wang}, X. {Wang}, Y. N. {Wang}, Y. {Wang}, Y. F. {Wang}, Y. F. {Wang}, P. {Wei}, M. Z. {Wei}, H. {Wu}, K. F. {Wu}, X. B. {Wu}, Y. Z. {Wu}, X. Z. {Xing}, L. Z. {Xu}, X. Q. {Xu}, Y. {Xu}, T. S. {Yan}, D. H. {Yang}, H. F. {Yang}, H. Q. {Yang}, M. {Yang}, Z. Q. {Yao}, Y. {Yu}, H. {Yuan}, H. B. {Yuan}, H. L. {Yuan}, W. M. {Yuan}, C. {Zhai}, E. P. {Zhang}, H. W. {Zhang}, J. N. {Zhang}, L. P. {Zhang}, W. {Zhang}, Y. {Zhang}, Y. X. {Zhang}, Z. C. {Zhang}, M. {Zhao}, F. {Zhou}, X. {Zhou}, J. {Zhu}, Y. T. {Zhu}, S. C. {Zou}, F. {Zuo}, {The first data release (DR1) of the LAMOST regular survey}, {\it Research in Astronomy and Astrophysics\/} {\bf 15}, 1095-1124 (2015).

\bibitem{Planck+20}
{Planck Collaboration}, N. {Aghanim}, Y. {Akrami}, M.~{Ashdown}, J.~{Aumont}, C.~{Baccigalupi}, M.~{Ballardini}, A.~J.~{Banday}, R.~B.~{Barreiro}, N.~{Bartolo}, S.~{Basak}, R.~{Battye}, K.~{Benabed}, J. P.~{Bernard}, M.~{Bersanelli}, P.~{Bielewicz}, J.~J.~{Bock}, J.~R.~{Bond}, J.~{Borrill}, F.~R.~{Bouchet}, F.~{Boulanger}, M.~{Bucher}, C.~{Burigana}, R.~C.~{Butler}, E.~{Calabrese}, J. F.~{Cardoso}, J.~{Carron}, A.~{Challinor}, H.~C.~{Chiang}, J.~{Chluba}, L.~P.~L.~{Colombo}, C.~{Combet}, D.~{Contreras}, B.~P.~{Crill}, F.~{Cuttaia}, P.~{de Bernardis}, G.~{de Zotti}, J.~{Delabrouille}, J. M.~{Delouis}, E.~{Di Valentino}, J.~M.~{Diego}, O.~{Dor{\'e}}, M.~{Douspis}, A.~{Ducout}, X.~{Dupac}, S.~{Dusini}, G.~{Efstathiou}, F.~{Elsner}, T.~A.~{En{\ss}lin}, H.~K.~{Eriksen}, Y.~{Fantaye}, M.~{Farhang}, J.~{Fergusson}, R.~{Fernandez-Cobos}, F.~{Finelli}, F.~{Forastieri}, M.~{Frailis}, A.~A.~{Fraisse}, E.~{Franceschi}, A.~{Frolov}, S.~{Galeotta}, S.~{Galli}, K.~{Ganga}, R.~T.~{G{\'e}nova-Santos}, M.~{Gerbino}, T.~{Ghosh}, J.~{Gonz{\'a}lez-Nuevo}, K.~M.~{G{\'o}rski}, S.~{Gratton}, A.~{Gruppuso}, J.~E.~{Gudmundsson}, J.~{Hamann}, W.~{Handley}, F.~K.~{Hansen}, D.~{Herranz}, S.~R.~{Hildebrandt}, E.~{Hivon}, Z.~{Huang}, A.~H.~{Jaffe}, W.~C.~{Jones}, A.~{Karakci}, E.~{Keih{\"a}nen}, R.~{Keskitalo}, K.~{Kiiveri}, J.~{Kim}, T.~S.~{Kisner}, L.~{Knox}, N.~{Krachmalnicoff}, M.~{Kunz}, H.~{Kurki-Suonio}, G.~{Lagache}, J. M.~{Lamarre}, A.~{Lasenby}, M.~{Lattanzi}, C.~R.~{Lawrence}, M.~{Le Jeune}, P.~{Lemos}, J.~{Lesgourgues}, F.~{Levrier}, A.~{Lewis}, M.~{Liguori}, P.~B.~{Lilje}, M.~{Lilley}, V.~{Lindholm}, M.~{L{\'o}pez-Caniego}, P.~M.~{Lubin}, Y. Z.~{Ma}, J.~F.~{Mac{\'\i}as-P{\'e}rez}, G.~{Maggio}, D.~{Maino}, N.~{Mandolesi}, A.~{Mangilli}, A.~{Marcos-Caballero}, M.~{Maris}, P.~G.~{Martin}, M.~{Martinelli}, E.~{Mart{\'\i}nez-Gonz{\'a}lez}, S.~{Matarrese}, N.~{Mauri}, J.~D.~{McEwen}, P.~R.~{Meinhold}, A.~{Melchiorri}, A.~{Mennella}, M.~{Migliaccio}, M.~{Millea}, S.~{Mitra}, M. A.~{Miville-Desch{\^e}nes}, D.~{Molinari}, L.~{Montier}, G.~{Morgante}, A.~{Moss}, P.~{Natoli}, H.~U.~{N{\o}rgaard-Nielsen}, L.~{Pagano}, D.~{Paoletti}, B.~{Partridge}, G.~{Patanchon}, H.~V.~{Peiris}, F.~{Perrotta}, V.~{Pettorino}, F.~{Piacentini}, L.~{Polastri}, G.~{Polenta}, J. L.~{Puget}, J.~P.~{Rachen}, M.~{Reinecke}, M.~{Remazeilles}, A.~{Renzi}, G.~{Rocha}, C.~{Rosset}, G.~{Roudier}, J.~A.~{Rubi{\~n}o-Mart{\'\i}n}, B.~{Ruiz-Granados}, L.~{Salvati}, M.~{Sandri}, M.~{Savelainen}, D.~{Scott}, E.~P.~S.~{Shellard}, C.~{Sirignano}, G.~{Sirri}, L.~D.~{Spencer}, R.~{Sunyaev}, A. S.~{Suur-Uski}, J.~A.~{Tauber}, D.~{Tavagnacco}, M.~{Tenti}, L.~{Toffolatti}, M.~{Tomasi}, T.~{Trombetti}, L.~{Valenziano}, J.~{Valiviita}, B.~{Van Tent}, L.~{Vibert}, P.~{Vielva}, F.~{Villa}, N.~{Vittorio}, B.~D.~{Wandelt}, I.~K.~{Wehus}, M.~{White}, S.~D.~M.~{White}, A.~{Zacchei}, A.~{Zonca}, {Planck 2018 results. VI. Cosmological parameters},  {\it \aap\/} {\bf 641}, A6 (2020).

\bibitem{Hoffman+1996}
G. L.~{Hoffman}, E.~E.~{Salpeter}, B.~{Farhat}, T.~{Roos}, H.~{Williams}, G.~{Helou}, {Arecibo HI Mapping of a Large Sample of Dwarf Irregular Galaxies}, {\it \apjs\/} {\bf 105}, 269 (1996).

\bibitem{McNichols+2016}
A. T.~{McNichols}, Y. G.~{Teich}, E.~{Nims}, J. M.~{Cannon}, E. A.~K. {Adams}, E. Z.~{Bernstein-Cooper}, R.~{Giovanelli}, M. P.~{Haynes}, G. I.~G. {J{\'o}zsa}, K. B.~W.~{McQuinn}, J. J.~{Salzer}, E. D.~{Skillman}, S. R.~{Warren}, A.~{Dolphin}, E.~C.~{Elson}, N.~{Haurberg}, J.~{Ott}, A.~{Saintonge}, I.~{Cave}, C.~{Hagen}, S.~{Huang}, S.~{Janowiecki}, M. V.~{Marshall}, C. M.~{Thomann}, A. {Van Sistine}, {SHIELD: Neutral Gas Kinematics and Dynamics}, {\it \apj\/} {\bf 832}, 89 (2016).

\bibitem{Bernstein-Cooper+2014}
E. Z. {Bernstein-Cooper}, J. M.~{Cannon}, E. C.~{Elson}, S. R.~{Warren}, J. {Chengular}, E. D.~{Skillman}, E. A.~K. {Adams},  A. D. {Bolatto}, R. {Giovanelli}, M. P. {Haynes}, K. B.~W.  {McQuinn}, S. A. {Pardy}, K. L. {Rhode}, J. J. {Salzer}, {ALFALFA Discovery of the Nearby Gas-rich Dwarf Galaxy Leo P. V. Neutral Gas Dynamics and Kinematics}, {\it \aj\/} {\bf 148}, 35 (2014).
\end{thebibliography}
\end{document}